\documentclass[twoside,11pt]{article}

\usepackage[hyphens]{url}

\usepackage[abbrvbib, accepted, specialissue]{melba}

\usepackage{amsmath,amsfonts}
\usepackage{caption}
\usepackage{subcaption}
\usepackage{multirow}
\usepackage{graphicx}
\usepackage{graphbox}
\usepackage{enumitem}
\usepackage{array}
\usepackage{hypcap}
\newcolumntype{P}[1]{>{\centering\arraybackslash}p{#1}}

\usepackage{tcolorbox}

\melbaid{2024:003}  % This is provided upon by the publishing editor
\doi{https://doi.org/10.59275/j.melba.2024-b87a}
\melbaauthors{Hassanaly, Brianceau, Solal, Colliot and Burgos}  % Note: this one is also used to set the pdf 'authors' metadata
\volume{2}
\firstpageno{611}
\melbayear{2024}  % The publication year
\datesubmitted{08/2023}  % Date submitted to MELBA: mm/yyyy
\datepublished{01/2024}  % Today's date: mm/yyyy

\melbaspecialissue{Special Issue for Generative Models}
\melbaspecialissueeditors{Mert Sabuncu, Sotirios A. Tsaftaris}

\ShortHeadings{Evaluation of pseudo-healthy image reconstruction for anomaly detection}{Hassanaly, Brianceau, Solal, Colliot and Burgos}

\title{Evaluation of pseudo-healthy image reconstruction for anomaly detection with deep generative models:\\ Application to brain FDG PET}

\author{\firstname Ravi Hassanaly\orcid{0009-0009-1304-5906} \email ravi.hassanaly@icm-institute.org \\  % start right after \author{, or there will be an extra space
	\addr Sorbonne Université, Institut du Cerveau - Paris Brain Institute - ICM, CNRS, Inria, Inserm, AP-HP, Hôpital de la Pitié Salpêtrière, F-75013, Paris, France
	\AND
	\name Camille Brianceau \email camille.brianceau@icm-institute.org \\
	\addr Sorbonne Université, Institut du Cerveau - Paris Brain Institute - ICM, CNRS, Inria, Inserm, AP-HP, Hôpital de la Pitié Salpêtrière, F-75013, Paris, France
        \AND
        \name Maëlys Solal\orcid{0000-0001-9333-9238} \email maelys.solal@icm-institute.org \\
	\addr Sorbonne Université, Institut du Cerveau - Paris Brain Institute - ICM, CNRS, Inria, Inserm, AP-HP, Hôpital de la Pitié Salpêtrière, F-75013, Paris, France
        \AND
	\name Olivier Colliot\orcid{0000-0002-9836-654X} \email olivier.colliot@cnrs.fr \\
	\addr Sorbonne Université, Institut du Cerveau - Paris Brain Institute - ICM, CNRS, Inria, Inserm, AP-HP, Hôpital de la Pitié Salpêtrière, F-75013, Paris, France
        \AND
	\name Ninon Burgos\orcid{0000-0002-4668-2006} \email ninon.burgos@cnrs.fr \\
	\addr Sorbonne Université, Institut du Cerveau - Paris Brain Institute - ICM, CNRS, Inria, Inserm, AP-HP, Hôpital de la Pitié Salpêtrière, F-75013, Paris, France
}

\begin{document}

% top matter
\maketitle

% abstract
\begin{abstract}%   <- trailing '%' for backward compatibility of .sty file
	Over the past years, pseudo-healthy reconstruction for unsupervised anomaly detection has gained in popularity. This approach has the great advantage of not requiring tedious pixel-wise data annotation and offers possibility to generalize to any kind of anomalies, including that corresponding to rare diseases. By training a deep generative model with only images from healthy subjects, the model will learn to reconstruct pseudo-healthy images. This pseudo-healthy reconstruction is then compared to the input to detect and localize anomalies. The evaluation of such methods often relies on a ground truth lesion mask that is available for test data, which may not exist depending on the application. 

    We propose an evaluation procedure based on the simulation of realistic abnormal images to validate pseudo-healthy reconstruction methods when no ground truth is available. This allows us to extensively test generative models on different kinds of anomalies and measuring their performance using the pair of normal and abnormal images corresponding to the same subject. It can be used as a preliminary automatic step to validate the capacity of a generative model to reconstruct pseudo-healthy images, before a more advanced validation step that would require clinician's expertise. We apply this framework to the reconstruction of 3D brain FDG PET using a convolutional variational autoencoder with the aim to detect as early as possible the neurodegeneration markers that are specific to dementia such as Alzheimer's disease.

\end{abstract}

% keywords

\section{Introduction}

During the last decade, breakthroughs in deep learning and computer vision combined with the increasing quality and number of medical data available have offered many new possibilities in medical image processing and analysis \citep{Litjens2017, esteva2017dermatologist}. Deep learning aims to become a new standard for analyzing medical images and algorithms are trained to accomplish tasks that require a high level of expertise, such as anomaly detection for diagnostic support \citep{Litjens2017, burgos2021deep, suganyadevi2022review}. A strategy for anomaly detection with deep learning consists in using a supervised algorithm that learns from annotated data. This has the advantage of having remarkable performance on the specific task learned, which can be classification between normal and abnormal images \citep{esteva2017dermatologist, wen_convolutional_2020} or anomaly segmentation \citep{zhou2018unet, isensee2021nnu}. However, this strategy has several drawbacks: the first one is that it requires a large amount of annotated data that are time consuming and costly to acquire. The second one is that the model's results will be affected by potential annotation errors. The last disadvantage is that the models will be specific to the data, diseases and anomalies they have been trained on. This might be an issue especially for rare diseases for which few data samples are available.

Another strategy is to use self-supervised, weakly supervised or unsupervised learning for anomaly detection \citep{chen2022unsupervised, zhang2023dive}. The underlying idea of these methods is to learn the distribution of healthy data. One can then use it to detect out-of-distribution samples, and thus identify abnormal cases. Another way is to use generative models to reconstruct pseudo-healthy images from the healthy data distribution: since the model is trained to reconstruct only normal data, we assume that the reconstruction of abnormal images will be imperfect, and by comparing the input real image to the reconstruction, we should be able to detect anomalies. The first advantage of this strategy is that it does not require voxel-level annotation. Another benefit is that it should be able to detect any type of anomaly, potentially linked to different diseases. Deep generative models such as variational autoencoders (VAEs) \citep{kingma2013auto}, generative adversarial networks (GANs) \citep{goodfellow_generative_2014} and more recently denoising diffusion probabilistic models (DDPMs) \citep{ho2020denoising} have shown great results for image generation tasks and unsupervised anomaly detection (UAD) in medical imaging \citep{esmaeili2023generative}, including neuroimaging \citep{wang2023applications, gong2023generative}.

Our work focuses on dementia, and more precisely on Alzheimer's disease (AD), its most common form. Therapies are increasingly becoming available. Very recently, several phase 3 trials of $\beta$-amyloid depleting therapies have demonstrated their effectiveness to slow down the progression of cognitive decline. This has led to approval of several treatments by the FDA either through the accelerated approval~\footnote{\url{https://www.fda.gov/news-events/press-announcements/fda-grants-accelerated-approval-alzheimers-drug}} or traditional pathway \footnote{\url{https://www.fda.gov/news-events/press-announcements/fda-converts-novel-alzheimers-disease-treatment-traditional-approval}} and another treatment has been submitted for approval. In each of these cases, it is essential to detect the disease as early as possible, and if possible before the appearance of the first symptoms, to ensure effectiveness of the medication. Medical imaging plays a key role to observe the physiological changes that appear in the brain several years before the symptoms, such as neurodegeneration, $\beta$-amyloid aggregation or accumulation of tau protein \citep{jackUnbiasedDescriptiveClassification2016}.
Positron emission tomography (PET) is a relevant imaging modality for Alzheimer's disease and dementia \citep{herholz_fdg_1995, herholz_positron_2007, quigley_pet_2011} as it allows imaging these three phenomena. PET images are 3D images that highlight the concentration of a radioactive tracer administered intravenously to the patient. Several tracers are commonly used for the diagnosis of Alzheimer's disease \citep{nordberg_use_2010}. Here we focus on \textsuperscript{18}F-fluorodeoxyglucose (FDG), which is a glucose analogue that is used to localize brain areas with altered glucose metabolism, hypometabolism in the case of neurodegeneration \citep{herholz_fdg_1995}.

In clinical practice, PET scans are mostly analyzed visually by a nuclear physician and reading interpretation highly depends on the physician expertise. We would like to automatize the detection of anomalies using pseudo-healthy reconstruction to provide the clinician a robust computer-aided diagnosis tool. UAD has been widely applied to neuroimaging data to detect anomalies that are sharp and with distinct outlines, such as tumors or white matter hyper-intensities on structural magnetic resonance images (MRI) \citep{baur_autoencoders_2021}. However, applying UAD to dementia is less common \citep{choi_deep_2019, baydargil_anomaly_2021} because the anomalies, e.g., metabolic changes on FDG PET, are more diffuse and less intense, making the task more challenging. Moreover, when applying UAD in the context of dementia, we do not have access to ground truth masks of the anomalies to evaluate the models, contrary to applications that aim to detect tumors or white matter hyper-intensities.

In this paper, we introduce a framework for the evaluation of pseudo-healthy reconstruction approaches in the absence of ground truth. This framework consists in simulating anomalies on images of healthy subjects to generate pairs of pathology-free and pathological (e.g., mimicking dementia-like lesions) images. We complement the framework by defining new healthiness and anomaly metrics. The healthiness metric measures whether the reconstructed image is of healthy appearance to evaluate the model capacity to reconstruct pseudo-healthy images, whereas the anomaly metric measures whether the input image contains anomalies using both the pseudo-healthy reconstruction and the input image. A preliminary version of this work was published as a conference paper~\cite{hassanaly2023simulation}. The present article extends the previous work mainly with:
(i) a more comprehensive motivation and methodological description of our evaluation strategy; 
(ii) the introduction of new metrics to complete the evaluation framework;
(iii) numerous experiments to evaluate a 3D VAE trained on full resolution PET using the framework, including analyses of the VAE latent space.

\section{Related Work}
\label{sec:related_work}

Unsupervised methods having many benefits, their use for anomaly and outlier detection in medical imaging has increased over the last years, especially methods based on image synthesis.

A family of methods do not rely on the image reconstruction but rather consist of out-of-distribution detection algorithms. For instance, \cite{alaverdyan2020regularized} used a siamese autoencoder on 3D patches to detect epilepsy lesions. The model learns a latent representation of the normal healthy data and a one-class SVM classifies abnormalities as outliers in the latent space. This method had further been improved and applied to Parkinson's disease \citep{pinon2023brain} and the detection of white matter hyper-intensities \citep{pinon2023oneclass}.

Concerning pseudo-healthy reconstruction, the underlying idea is to create new images from existing data using a generative model such a VAE \citep{kingma2013auto}, GAN \citep{goodfellow_generative_2014} or DDPM \citep{ho2020denoising}. By training only with images from healthy subjects, the model learns the distribution of normal or healthy data. The reconstruction of an image with this model will look like a healthy version of the original image, whether the image is that of a healthy subject or a patient with a disease, thus the name pseudo-healthy reconstruction. The synthesized pseudo-healthy image is finally compared to the real one to detect anomalies and possibly compute an anomaly score.

Pseudo-healthy reconstruction has been used in numerous fields of medical imaging \citep{fernando_deep_2022}, for instance to detect various lung anomalies such a pneumonia on chest x-ray \citep{nakao2021unsupervised, kim2023unsupervised}, retinal anomalies on optical coherence tomography \citep{schlegl_unsupervised_2017, schlegl_f-anogan_2019, zhou2023spatial}, breast cancer on mammogram \citep{park2023unsupervised}, skin cancer on dermatoscopic images \citep{lu2018anomaly}, tumors detection on PET, computed tomography (CT) and PET-CT \citep{astaraki2023autopaint}, or malignant tissues on colonoscopy \citep{tian2021constrained}. In the following, we will focus on methods applied to neuroimaging.

Several methods based on autoencoders have been developed, starting by \cite{zimmerer_context-encoding_2018} who used a context-encoding VAE for the detection of brain glioma and multiple sclerosis lesions on anatomical MRI. The localization of the anomalies was improved in a subsequent work using a pixel-wise KL distance \citep{zimmerer_unsupervised_2019}. Another work by \cite{chen_unsupervised_2018} introduced constrained adversarial autoencoders. \cite{marimont2021anomaly} used both prior-based anomaly score and reconstruction-based anomaly score with a vector quantized VAE (VQ-VAE). \cite{pinaya2022unsupervised} also used a VQ-VAE together with an autoregressive transformer in the latent space to better learn the probability density function of healthy data. To show the efficiency of autoencoders, \cite{baur2021modeling} implemented an autoencoder with a spatial latent space and skip-connections and compared the result to a UNet trained for supervised anomaly segmentation. \cite{bercea2023generalizing} tried to generalize UAD to non hyper-intense anomalies in order to detect various pathological features using a reverse autoencoder. \cite{luth2023cradl} reused the general principle of pseudo-healthy reconstruction with autoencoders but in their case, the encoder is improved thanks to contrastive learning in order to use high level features of the image to learn a better latent representation.

Based on the fundamental work of \cite{schlegl_unsupervised_2017} who introduced AnoGAN and its improved version, the f-AnoGAN \citep{schlegl_f-anogan_2019}, several frameworks using GANs have been developed, for instance VAE-GAN \citep{baur_deep_2019}, ANT-GAN \citep{sun_adversarial_2020}, or cycleGAN \citep{xia_adversarial_2019, xia_pseudo-healthy_2020}. More recently, \cite{shi2023generative} introduced GANCMLAE, a GAN-based approach combined with an autoencoder and constrained by multiple losses for the early detection of brain atrophy. Another novel and interesting approach has been proposed by \cite{siddiquee2023brainomaly}, who train a GAN-based model with both healthy and abnormal images to have a fully unsupervised method. Finally, \cite{bercea2023reversing} combined both a latent generative model and high quality reconstruction networks based on GAN to detect stroke lesions on T1-weighted MRI. \cite{baur_autoencoders_2021} summed up and compared many of the VAE and GAN approaches that had been used for unsupervised brain tumor and multiple sclerosis lesion segmentation on MRI data.

More recently, following the success of diffusion models for image generation, DDPMs have also been used for anomaly detection tasks in medical imaging \citep{wolleb2022diffusion, pinaya2022fast, bercea2023mask}.

Most methods for UAD have been applied to brain structural MRI, often targeting sharp and visible anomalies such as tumors or multiple sclerosis lesions. Only a few studies have focused on other modalities such as computed tomography or PET, probably because less data is available. 
\cite{choi_deep_2019} implemented a simple VAE for anomaly detection on brain FDG PET. \cite{baydargil_anomaly_2021} used a GAN with an autoencoder architecture for the generator (with a parallel model for the encoder) to detect anomalies on FDG PET in the context of Alzheimer's disease. 

In most cases, the proposed methods work with 2D images that are extracted from 3D volumes. But recently, numerous articles working directly with 3D images or trying to reconstruct 3D volumes have been published. \cite{pinaya2022unsupervised} validated their model on both 2D and 3D images. \cite{chatterjee2022strega} proposed a compact version of the context encoding VAE of \cite{zimmerer_context-encoding_2018} that is trained on 2D slices that are stacked to obtain a 3D volume. \cite{han2021madgan} presented a similar strategy, which consists of using three successive slices to reconstruct the following three slices to take into account the 3D structure of the image. \cite{luo2023unsupervised} directly trained a 3D encoder to detect brain abnormalities on T2-weighted volumes. \cite{bengs2021three} compared 3D and 2D VAEs for anomaly detection on brain MRI. \cite{bengs2022unsupervised} trained a VAE on 3D T1-weighted MRI by additionally considering the age information. \cite{simarro2020unsupervised} proposed a 3D extension of the 2D f-AnoGAN and refined the training steps to detect traumatic brain injuries.

Many studies have used the BraTS (glioma) \citep{menze2014multimodal}, ISLES (multiple sclerosis lesions) \citep{maier2017isles} or ATLAS (stroke lesions) \citep{liew2017anatomical} datasets, which directly provide ground truth anomaly masks \citep{zimmerer_context-encoding_2018, zimmerer_unsupervised_2019, chen_unsupervised_2018, baur_autoencoders_2021,  bercea2023generalizing, bercea2023reversing, luth2023cradl, wargnier2023weakly, pinaya2022unsupervised, chatterjee2022strega, luo2023unsupervised, bengs2022unsupervised, xia_adversarial_2019, xia_pseudo-healthy_2020, sun_adversarial_2020}. Other studies have used in-house data that may include ground truth anomaly masks \citep{baur_deep_2019, baur2021modeling, siddiquee2023brainomaly, alaverdyan2020regularized, luo2023unsupervised, han2021madgan}. In that case the evaluation of the model is straightforward: one only has to compute a metric such as the dice score between the predicted anomaly and the ground truth, as we would do for the evaluation of supervised anomaly segmentation. Some works have gone further by introducing new original metrics: \cite{xia_pseudo-healthy_2020} defined a "healthiness" metric, using a segmentation network to estimate the size of a potential lesion in the pseudo-healthy reconstruction; and an ``identity" metric, based on a multi-scale structural similarity index on non pathological tissues. In most of the other cases, when the ground truth anomaly mask is not available, the evaluation consists in applying a classifier to the reconstructed images that was trained to distinguish pathological and healthy images, or using the reconstruction error itself from which an anomaly score is derived. One way to improve the evaluation is to use synthetic data by corrupting real healthy data with sprites \citep{bercea2023generalizing, pinaya2022unsupervised}.

Strategies developed for pseudo-healthy reconstruction and, more generally, unsupervised anomaly detection, often lack rigorous evaluation. Furthermore, the majority of studies utilize 2D images, with very few focusing on PET images. This is why, in this paper, we introduce an evaluation framework particularly adapted for experiments where ground truth data is unavailable. Subsequently, we apply this framework to conduct a rigorous evaluation of a 3D model for pseudo-healthy reconstruction. We apply this model to the detection of anomalies associated with dementia, a task that has received limited exploration and presents significant challenges.

\section{Methods}
\label{sec:methods}

\subsection{Variational autoencoder for pseudo-healthy image reconstruction}

We opt for a 3D convolutional VAE as VAEs have already shown their efficacy for UAD in medical imaging \citep{baur_autoencoders_2021,chen2022unsupervised}: they are easy to train, easily scalable, are able to handle small datasets, and with good interpretation capacity thanks to their regularized latent space. Moreover the VAE framework allows us to really learn the training data distribution which is an import point specifically for our study as we will see later. We implement a 3D VAE to fully exploit the 3D context of high resolution PET images.

A VAE is a deep learning model \citep{kingma2013auto} combining two parameterized models: the encoder (or recognition model) that will map the dataset unknown distribution to a known prior distribution, and the decoder (or generative model) that will generate data from latent samples that are likely to be in the dataset distribution.

Let's consider $D$ a set of medical images $\mathbf{x}$. Our goal is to approximate the unknown true distribution of the data $p(\mathbf{x})$ with a parameterized model $p_{\theta}(\mathbf{x})$, by learning a set of parameters $\theta$. To do so, we use a deep latent variable model that will allow us to map the complex unknown true distribution $p(\mathbf{x})$ to a latent distribution $p(\mathbf{z})$ that is selected.
Let $\mathbf{z}$  be a random vector jointly-distributed with $\mathbf{x}$. We assume that $\mathbf{z}$ is involved in the generation process of $\mathbf{x}$

\begin{equation*}
    p_{\theta}(\mathbf{x}) = \int\displaylimits_z p_{\theta}(\mathbf{z})p_{\theta}(\mathbf{x\mid z})\mathrm{d}z \enspace ,
\end{equation*}

where $p_{\theta}(\mathbf{z})$ is the known prior distribution and $p_{\theta}(\mathbf{x\mid z})$ is the generative model, also called decoder.

To compute the appropriate latent representation $\mathbf{z}$ for each image $\mathbf{x}$ of our dataset, we need the posterior distribution $p_{\theta}(\mathbf{z\mid x})$. However, as the posterior $p_{\theta }(\mathbf{z} \mid \mathbf{x})$ is intractable, it is necessary to introduce a parametric inference model $q_{\Phi}(\mathbf{z\mid x})$ to approximate the posterior distribution: $q_{\Phi }(\mathbf{z\mid x})\approx p_{\theta}(\mathbf{z\mid x})$ with $\Phi$ parameterizing $q$. The model $q_{\Phi}(\mathbf{z\mid x})$ is called the recognition model or encoder.

Finally we have two models $q_{\Phi}(\mathbf{z\mid x})$ and $p_{\theta}(\mathbf{x\mid z})$ (the encoder and the decoder) with two sets of parameters $\Phi$ and $\theta$ (the weights of the neural networks) to optimize.

In order to train the model and find optimums for $\Phi$ and $\theta$, we will use the observations $\mathbf{x}$ of our dataset $D$ to maximize the log-likelihood of our model. \cite{kingma2013auto} showed that this is equivalent to maximizing the evidence lower bound, which defines our loss function $\mathcal{L}_{\theta, \Phi}$ 

\begin{equation}
    \log{(p_{\theta}(\mathbf{x}))} \geq \mathcal{L}_{\theta, \Phi}(\mathbf{x}) = \mathbb{E}_{q_{\Phi}\mathbf{(z\mid x})} \Bigl[ \log  \bigl( p_{\theta}(\mathbf{x\mid z}) \bigr) \Bigr] - D_{{\mathrm{KL}}} \Bigl( q_{\Phi}(\mathbf{z\mid x})\|p_{\theta}(\mathbf{z}) \Bigr) \enspace ,
\label{eq:loss}
\end{equation}
with $D_{\mathrm{KL}}$ the Kullback-Leibler divergence.
The idea is to jointly optimize the set of parameters $\theta$ to improve the generated data quality, which means minimizing the reconstruction error, and the set of parameters $\Phi$ such that $q_{\Phi}(\mathbf{z\mid x})$ is the closest to posterior $p_{\theta}(\mathbf{z\mid x})$.

We approximate all the distributions by Gaussian distributions: we have the prior $p_{\theta}(\mathbf{z}) = \mathcal{N} (0, I)$, and the posterior $q_{\Phi}(\mathbf{z\mid x}) = \mathcal{N} (\mu(\mathbf{x}), \sigma(\mathbf{x}))$ with $\mu(\mathbf{x})$ and $\sigma(\mathbf{x})$ being respectively the mean and the standard deviation of the probability distribution of a true sample $\mathbf{x}$ in the latent space. This gives us the following expression:
\begin{equation*}
\begin{split}
    D_{{\mathrm{KL}}}(q_{\Phi}(\mathbf{z\mid x})\|p_{\theta}(\mathbf{z})) & =  \frac{1}{2} \left[ 1 + \log(\sigma(\mathbf{x})^{2}) - \sigma(\mathbf{x})^{2} - \mu(\mathbf{x})^{2} \right]  \enspace .
\end{split}
\end{equation*}

Finally, we approximate $p_{\theta}(\mathbf{x\mid z})$ with a normal distribution such that  $p_{\theta}(\mathbf{x\mid z}) = \mathcal{N} (f(\mathbf{z}), cI)$, with $f(\mathbf{z})$ being the decoder output of a latent vector $\mathbf{z}$, and $cI$ being a covariance matrix with $c\in\mathbb{R}$. The reconstruction term $\mathbb{E}_{q_{\Phi}\mathbf{(z\mid x})} \Bigl[ \log  \bigl( p_{\theta}(\mathbf{x\mid z}) \bigr) \Bigr]$ is proportional to the mean squared error (MSE) between the input of the VAE $\mathbf{x}$ and the reconstruction $\hat{\mathbf{x}} = f(\mathbf{z})$: $\mathbb{E}_{q_{\Phi}\mathbf{(z\mid x})} \Bigl[ \log  \bigl( p_{\theta}(\mathbf{x\mid z}) \bigr) \Bigr] \propto MSE(\mathbf{x}, f(\mathbf{z}))$ . This gives us the final loss that we will use
\begin{equation}
\begin{split}
    \mathcal{L}_{\theta, \Phi}(\mathbf{x}) = MSE(\mathbf{x}, \hat{\mathbf{x}}) - \frac{1}{2} \left[ \sigma(\mathbf{x})^{2} + \mu(\mathbf{x})^{2} - \log(\sigma(\mathbf{x})^{2}) - 1 \right] \enspace .
\end{split}
\end{equation}

In our pseudo-healthy image synthesis use case, $D$ is a dataset containing images from healthy subjects and patients, and can be divided into respectively two complementary subsets $D_{h}$ and $D_{p}$. As we want to learn the distribution of healthy images $D_{h}$, the model is trained using only images from $D_{h}$. Then, when reconstructing an image during testing from $\mathbf{x}\in D$, the reconstruction $\mathbf{\hat{x}}$ should be in $D_{h}$. Indeed, if $\mathbf{x_h}\in D_{h}$, then the reconstruction should be almost equal to the input: $\mathbf{x_h}\approx\mathbf{\widehat{x_h}}$. However, if the input image is pathological (i.e., with anomalies) $\mathbf{x_p}\in D_{p}$, then we want the reconstruction to be pseudo-healthy, which means that we want $\mathbf{\widehat{x_p}}\in D_{h}$. Since the decoder has been trained to generate healthy images, for each latent representation $\mathbf{z}$, the reconstruction $p_{\theta}(\mathbf{x}\mid \mathbf{z})$ is likely to be in the distribution of $D_h$, whether $\mathbf{z}$ is a latent representation of a healthy image $\mathbf{x_h}\in D_h$ or a pathological one $\mathbf{x_p}\in D_p$. We also have to make sure that the reconstruction is subject-specific, which translates into the fact that the encoder can predict a unique latent representation $\mathbf{z}$ for a subject independently of its diagnosis. In mathematical terms, we can write it as $q_{\Phi}(\mathbf{z} \mid \mathbf{x_h}) \approx q_{\Phi}(\mathbf{z} \mid \mathbf{x_p})$ with $\mathbf{x_h}$ and $\mathbf{x_p}$ being two images of the same subject. This hypothesis will be verified in the Section~\ref{sec:method_latent}.

\subsection{Pseudo-healthy image reconstruction evaluation procedure}

Rigorous and in-depth evaluation of machine learning models and of their training procedure is crucial, especially in the medical field as overestimated or biased results may lead to dramatic consequences \citep{varoquaux2022evaluating}. As far as we know, there is no guidelines nor standard procedure for the evaluation of pseudo-healthy reconstruction for UAD, especially when a ground truth of the anomalies that should be detected is not available. We propose here such procedure.

In this context, we can identify two objectives: 
i) preserve the identity of the subject in the reconstructed image, ii) reconstruct an image of healthy appearance \citep{xia_pseudo-healthy_2020}. We have to evaluate the performance of the model for both objectives.
For the first one, we can measure the similarity between images of healthy subjects and their reconstruction.
With regard to the second objective, we can either evaluate whether the images reconstructed are looking healthy (pseudo-healthy reconstruction task), or measure if the anomalies detected by this method correspond to the real anomalies present in the image (anomaly detection task). However, depending on the type of disorder studied, we may not have ground truth healthy images, nor ground truth anomaly masks, so we cannot use a metric to quantify how healthy the reconstructed images are nor how well anomalies are detected. This is why we developed an evaluation framework that consists in simulating an abnormal image $\mathbf{x'}$ from a healthy image $\mathbf{x}$ in order to have a pair with a diseased image and its healthy version. To evaluate the healthiness of a reconstructed image $\mathbf{\widehat{x'}}$ from an abnormal simulated image $\mathbf{x'}$, we can measure the similarity between the pseudo-healthy reconstruction $\mathbf{\widehat{x'}}$ and the the original healthy image $\mathbf{x}$.

\subsubsection{Evaluation metrics for image reconstruction}

The first step to validate a pseudo-healthy reconstruction model is to evaluate %the preservation of the subject's identity, which is equivalent to evaluating the 
the quality of the reconstruction in the case of images of healthy subjects. We use four metrics that are common in the image synthesis literature \citep{Necasova2022ValidationEvaluation}: the mean squared error (MSE), the peak signal-to-noise ratio (PSNR), the structural similarity index (SSIM) \citep{wang2004image} and the multi-scale structural similarity (MS-SSIM) \citep{wang2003multiscale} (see details in Appendix~\ref{sec:recon_metrics}). This also aims validating the fact that the model can reconstruct images that are as healthy as they originally look. 

However, since the reconstruction metrics are computed on the whole 3D image and not only in the abnormal region, their values do not substantially vary when computed for healthy or abnormal images, as the major part of the image is normal. This is amplified when the anomalies are subtle. Thus we cannot rely only on whole image reconstruction metrics to differentiate healthy subjects from patients.

\subsubsection{Simulation-based evaluation framework}
\label{sec:eval_framework}

In practice, the healthy version of an image with anomalies is rarely available, it is thus impossible to measure the healthiness of the reconstructed image. In most of the studies on UAD in medical imaging, the datasets provide lesion masks that can be used as ground truths. In this case, one can simply compute a metric such as the dice score between the anomaly map generated (and usually post-processed to binarize it) and the real lesion mask to evaluate the capacity of the model to detect and localize anomalies. This can be used as a proxy measure of the healthiness of the reconstructed image: if we perfectly detect lesions, it means that the reconstruction is healthy compared to the input image. However, when studying disorders such as dementia, such lesion masks are not available.

Another way to evaluate the healthiness of the reconstructed images would be to consult a neuro-radiologist or nuclear physician. However, manually rating images is time consuming, especially if the aim is to compare different models, and possible only for small datasets. We are thus looking for a strategy to automatically evaluate models, as a preliminary validation before soliciting clinicians.

The idea is to simulate abnormal images to evaluate our model. In the literature, anomalies are often under the form of sprites (i.e. non realistic artifacts added to the images) \citep{bercea2023generalizing, pinaya2022unsupervised}. However this is not satisfactory as we try to detect subtle and less intense anomalies. Realistic anomaly generation has also been explored, mainly to study the progression of diseases such as cancer or dementia \citep{manzanera2021patient}. When reducing the scope to neuroimaging, most anomaly generation methods are applied to structural MRI; for instance to simulate the growth of a glioblastoma \citep{ezhov2023learn}, or the progression of atrophy in case of dementia \citep{khanal2017simulating, ravi2022degenerative}. The proposed approaches often rely on complex modeling or the use of deep learning.

\begin{figure}[!htb]
    \centering
    \includegraphics[width=\textwidth]{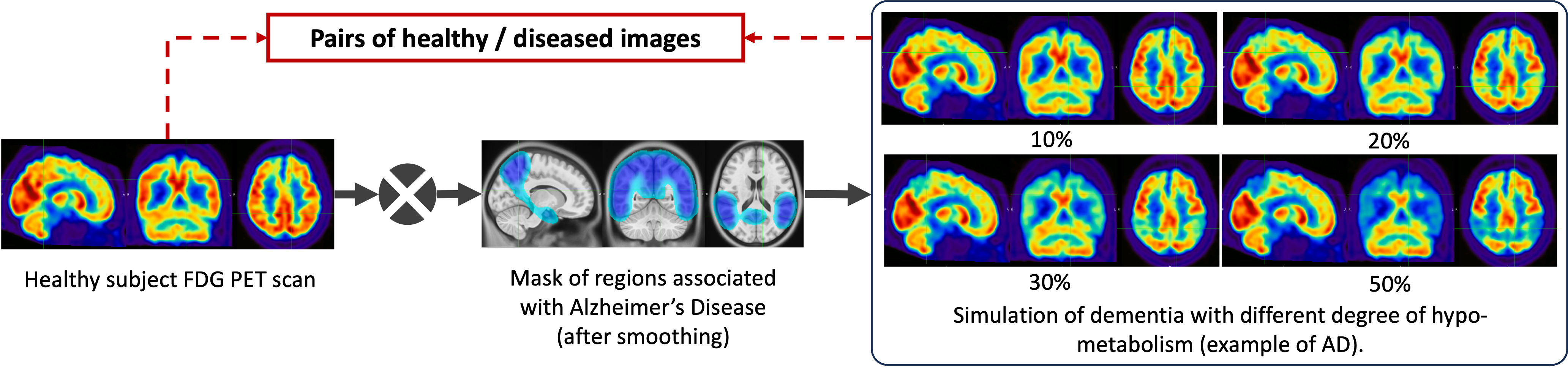}
    \caption{
    Hypometabolism simulation pipeline. The intensity of the image from a healthy subject is reduced by a chosen factor in a region associated with a dementia.
    }
    \label{fig:simulation_pipeline}
\end{figure}

\begin{figure}[!htb]
    \centering
    \includegraphics[width=0.75\textwidth]{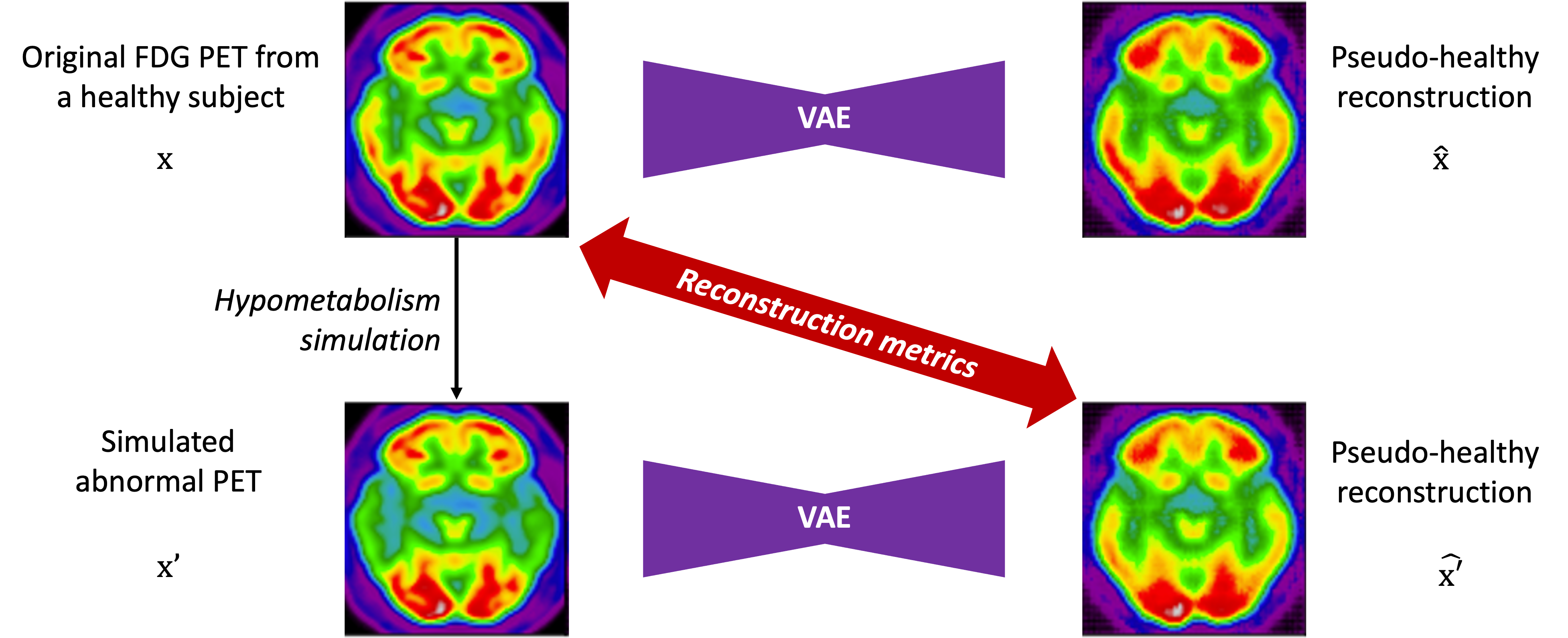}
    \caption{
    Evaluation framework using simulated images. We simulate an abnormal PET scan $\mathbf{x'}$ from an image of a healthy subject $\mathbf{x}$. If the model works perfectly, the reconstruction $\mathbf{\widehat{x'}}$ should be identical to the original image $\mathbf{x}$.
    }
    \label{fig:evalmethode}
\end{figure}

We here propose to simply generate new test sets by simulating hypometabolism on healthy images to have pairs of healthy (considered as ground truth) and abnormal images. For this purpose, we designed a mask corresponding to regions associated with AD (parietal and temporal lobes) \citep{Landau2012AmyloidDeposition} that were extracted from the third automated anatomical labelling (AAL3) atlas \citep{rolls2020automated}. To obtain a realistic simulated image, we smoothed the mask with a Gaussian convolution filter with $\sigma = 5$. We then reduced the intensity of the PET signal within the region defined by the mask by different factors to simulate various degrees of hypometabolism as illustrated in Figure~\ref{fig:simulation_pipeline}. Having such pairs of images allows us to compare the pseudo-healthy image reconstructed by the model $\mathbf{\widehat{x'}}$ from images presenting anomalies $\mathbf{x'}$ with their corresponding healthy images $\mathbf{x}$ (Figure~\ref{fig:evalmethode}), hence better evaluating the model capacity to synthesize pseudo-healthy images.

To ensure that the UAD model being evaluated can generalize to dementia other than AD, we also generated masks corresponding to five other dementia subtypes: behavioral variant frontotemporal dementia (bvFTD), logopenic variant primary progressive aphasia (lvPPA), semantic variant PPA (svPPA), nonfluent variant PPA (nfvPPA) and posterior cortical atrophy (PCA) based on the regions defined by \cite{Burgos2021AnomalyDetection}. All the details about the selected regions are available in Appendix~\ref{sup:similation_framework} and a pipeline to use the simulation framework has been integrated into the ClinicaDL open-source software \citep{thibeau-sutreClinicaDLOpensourceDeep2022}.

This framework will help us to extensively evaluate our model on different kinds of anomalies (different shapes, locations and intensities) and allow us to define a new metric to assess the healthiness of reconstructed images.

\subsubsection{Measuring the healthiness of reconstructed images}
\label{sec:healthiness}

Now that we have pairs of healthy and abnormal images, we want to define a metric that would help evaluate if the model is able to reconstruct images that are looking healthy. We call this metric "healthiness" and denote it as $H$. We can define $H$ as follows:

\begin{equation}
    H = \frac{\mu_M}{\mu_{\Bar{M}}} \enspace ,
\end{equation}
with $\mu_M$ the average uptake in the region of the mask $M$ used to simulate the anomaly and $\mu_{\Bar{M}}$ the average uptake of voxels in the brain excluding the mask $M$. 

This metric compares the average uptake in the region in which we simulate the disease and the other regions of the brain. For an image from a healthy subject $\mathbf{x}$, the average uptake in $M$ is similar to the one in $\Bar{M}$, so $H$ will be close to 1. However for a simulated image $\mathbf{x'}$, as the intensity is decreased within the mask $M$, $H$ will be lower than one. We then have to measure if the healthiness of the pseudo-healthy reconstruction $\mathbf{\widehat{x'}}$ is similar to the one of the original image $\mathbf{x}$ (close to 1), or at least superior to that of the input $\mathbf{x'}$. This can also be used to measure healthiness for hyper-intense anomalies: the score of the input image would then be above 1 and, similarly to hypometabolism detection, the reconstruction's score should be around 1 (or at least lower than the input image healthiness).

This simple metric will allow us to evaluate the performance of the model for reconstructing healthy images, but as it uses the framework described in Section~\ref{sec:eval_framework}, it cannot then be used on real images since we do not have the a priori information on the location of anomalies. This is why we introduce another method to detect anomalies with the pseudo-healthy reconstruction.

\subsubsection{Anomaly detection and localization}

The goal of this other method is to localize and assess the severity of anomalies in real images of patients by comparing them to their pseudo-healthy reconstruction. The idea is similar to the healthiness metric but we cannot use an anomaly mask to compute the metric. Instead, we use a brain atlas to define regions in which we compare the average uptake between the input image $\mathbf{x}$ and the pseudo-healthy reconstruction $\mathbf{\widehat{x'}}$. This region-wise anomaly score allows us to assess if an image contains anomalies and if so, to localize them.

This method can be validated using the evaluation framework described in Section~\ref{sec:eval_framework} as we know where the anomalies are.

We define the regions that we use starting from that of the second automated anatomical labelling (AAL2) atlas \citep{rolls2015implementation}. To simplify the analysis, we merged the 120 regions into 23 regions: orbitofrontal, dorsolateral prefrontal (DLPFC), ventromedial prefrontal (VMPFC), motor, opercular, medial temporal, lateral temporal, temporal pole, sensory, medial occipital, lateral occipital, medial parietal, lateral parietal anterior cingulate gyrus, middle cingulate gyrus, posterior cingulate gyrus, midbrain, amygdala, thalamus, insula, hippocampus, cerebellum and cerebellar vermis. We refine these regions using the gray matter mask of the MNI ICBM 2009c Nonlinear Symmetric template \citep{fonov_unbiased_2009, fonov_unbiased_2011} to keep only the tracer uptake in the gray matter.

\subsection{Learning the data distribution}
\label{sec:method_latent}

One of the main advantages of the VAE over other generative models is its consistent latent space that can help to explain and interpret the results obtained. We run several experiments on the latent space to verify that the VAE can learn the healthy image distribution since there are only images of healthy subjects in the training set. %At inference, when reconstructing a supposedly pseudo-healthy image, the input image is first projected through the encoder into the distribution of healthy images in the latent space. Then, when the decoder reconstructs the image from this latent representation, it is possible to assess whether the reconstructed image is pseudo-healthy. 

%To verify that the learned distribution corresponds to the distribution of healthy images, 
We first visualize the latent space using a principal component analysis (PCA) to reduce the latent dimension from 256 to 2. We fit the PCA on the latent representation of healthy images from the training set. Then we use this PCA to predict the principal components of images of the test sets (real ones and simulated ones). Even if we plot only the first two principal components, this already indicates how the encoder behaves. This will help us to verify whether the learned posterior is the same for healthy and abnormal images, i.e., whether $q_{\phi}(\mathbf{z\mid x_h}) \approx q_{\phi}(\mathbf{z\mid x_p})$.

We can also verify that if two images are close in the image space, there are close in the latent space and vice versa. We compare the distance between images in both latent- and image-space. More precisely, for each image latent vector $z_i$ of the dataset, we compute the Minkowski distance with the latent vectors $z_j$ of all the other images

\begin{equation}
    D_{Minkowski} =  \left( \sum_{i=1}^n \left| z_{i} - z_{j} \right|^{p} \right)^{\frac{1}{p}} \enspace .
\end{equation}

We arbitrarily choose $p=10$ in all our applications as we wanted $p$ to be high enough to compute distance in a space of dimension 256.

When studying dementia, datasets are often longitudinal, which means that several images are available per subject. This allows us to first evaluate an intra-subject distance that is computed between a certain image of a subject and all the other images available for the same subject. We can also compute an inter-subject distance. For a certain image of a subject, the closest image from another subject is selected as well as the tenth, the twentieth, the thirtieth and the fortieth closest images in the latent space (after discarding images from the same subject). We then compute the Euclidean $L_{2}$ norm and the SSIM between our image and the collection of five images selected. Computing these distances in both the latent space and the image space allows identifying potential correlation between the two representations. To this end, we fitted a linear mixed effect model (LMM) to estimate the tendency of the evolution of the distance in the latent space with regards to the distance in the image space (see Appendix~\ref{sec:lmm}).

\subsection{Materials}

FDG PET scans used in this study were obtained from the ADNI database \citep{Jagust2010AlzheimerDisease, Jagust2015AlzheimerDisease}. The ADNI was launched in 2003 as a public-private partnership, led by Principal Investigator Michael W. Weiner, MD. The primary goal of ADNI has been to test whether serial MRI, PET, other biological markers, and clinical and neuropsychological assessment can be combined to measure the progression of mild cognitive impairment and early AD.

We selected FDG PET images co-registered, averaged and uniformized to a resolution of 8 mm full width at half maximum to reduce the variability due to the use of different scanners. We only used PET images for which a T1-weighted (T1w) MR image was available at the same session for preprocessing purposes. The images were then processed using Clinica's \citep{clinica} \texttt{pet-linear} pipeline: they were registered using a rigid transformation to the corresponding T1w MRI of the same session, and then affinely registered to the MNI ICBM 2009c Nonlinear Symmetric template \citep{fonov_unbiased_2009, fonov_unbiased_2011} using the transformation computed with the \texttt{t1-linear} pipeline. They were then normalized in intensity using the average PET uptake in a region comprising cerebellum and pons, and cropped. In the end, the dimension of the PET scan is $169\times208\times179$ with 1~mm isotropic voxels.

To filter out potential PET images not well registered to the MNI template, we performed quality control. We first controlled the quality of the registration between the T1w MRI and the MNI template as it is an intermediate step when registering the PET image to the MNI space. The approach relies on a pre-trained neural network called DARQ \citep{fonov2022darq} that learned to classify images that are adequately registered to the MNI template. We then assessed the quality of the alignment of the PET image itself with the MNI template. Here the approach relies on a metric that measures the overlap between the output of the \texttt{pet-linear} pipeline, i.e. the PET image supposedly aligned with the MNI template, and a mask corresponding to the outside of the brain obtained from the MNI template. If the overlap is large, we assume that the PET image is not well-registered. Both quality control pipelines are available in the ClinicaDL open-source software \citep{thibeau-sutreClinicaDLOpensourceDeep2022}.

In the ADNI database, there is a total of 3511 FDG PET scans from 1600 participants. This includes 554 cognitively normal (CN) subjects (1010 images) that we selected since UAD models are trained only on images from healthy subjects. We know that physiological changes can appear several years before the first clinical symptoms, so to ensure that images really correspond to a healthy brain, we kept only scans from subjects that are CN for at least three years after the session considered. All the details about data selection and quality control are in Appendix~\ref{sec:data_select}. Finally, we have 739 images from 378 CN subjects and 353 images at baseline from 353 AD patients.

\subsection{Model training}

We split our dataset of 378 CN subjects into training, validation and test sets at the subject's level to avoid any form of data leakage \citep{wen_convolutional_2020}. The split is stratified by sex and age to reduce biases. We keep only baseline sessions in test and validation sets to avoid biased results. 60 CN subjects (60 images) compose the test set that is used to assess whether the healthy images are reconstructed as healthy. We then divide our training set into six folds to estimate the variance due to data splitting: 53 subjects (53 images) belong to the validation sets to monitor the training and 265 subjects are used to train our models. This represents between 510 and 538 images for the training phase depending on the fold. In addition to CN subjects, we use the images of 353 AD patients acquired at baseline to create a second test set. Finally, we use the 60 images from CN subjects in the test set to build new test sets by using our simulation method: we simulate hypometabolism using masks on the healthy images and keep it as separate test sets. We simulate different intensities of AD from 5\% to 70\% and simulate other dementia subtypes at 30\% hypometabolism which results in a total of 12 simulated test sets.

\begin{table}[!htb]
    \caption{Description of the population of our train and test sets from the ADNI database}
    \renewcommand{\arraystretch}{1.25}
    \centering
    \begin{tabular}{l|l|lll}
        Diagnosis           & Train/Test & N subjects & Age at baseline   & Sex (\%F/\%H) \\ \hline
        \multirow{2}{*}{CN} & Train      & 318        & 73.3 ± 6.1 & 51\%/49\%       \\
                            & Test       & 60         & 73.5 ± 6.6 & 63.3\%/36.7\%       \\ \hline
        AD                  & Test       & 353        & 75.3 ± 7.5 & 41.1\%/58.9\%      
    \end{tabular}
\end{table}

\begin{figure}[!htb]
    \centering
    \includegraphics[width=0.9\textwidth]{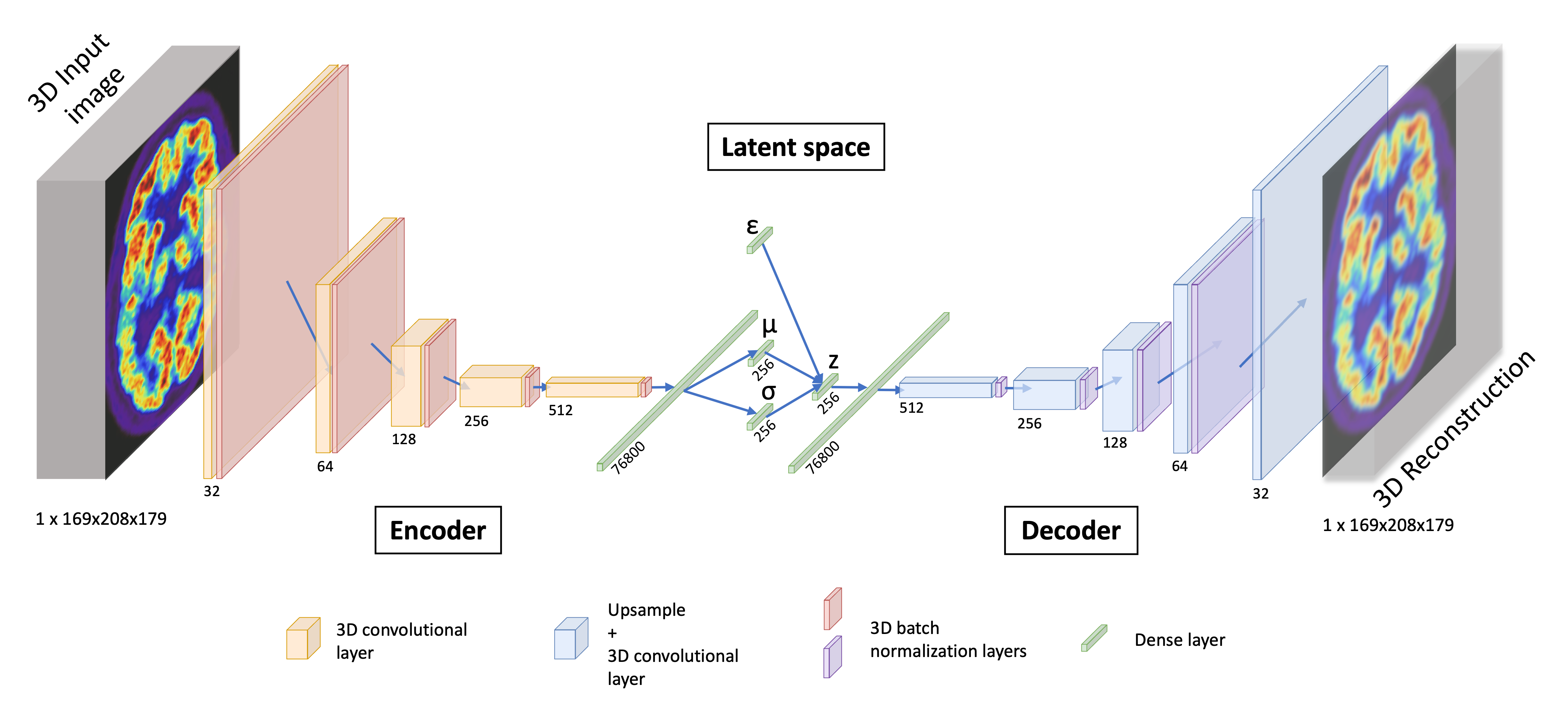}
    \caption{Architecture of the 3D convolutional VAE.}
    \label{fig:vae}
\end{figure}

Regarding the VAE implemented, the encoder is composed of five convolutional blocks that are the succession of a 3D convolutional layer and a batch normalization with a ReLU activation. Then the vector is flattened and passes through a dense layer to output the latent space of size 256 in one dimension. Our decoder is almost symmetrical as it transforms a single vector sampled from the latent space in a 3D image with a dense layer followed by four deconvolutional blocks that are composed of an upsampling layer, a 3D convolutional layer and a batch normalization with a leaky ReLU activation. The output /rev{block is composed of an upsampling layer and a 3D convolutional layer} with a sigmoid activation. Encoder convolutional layers have a kernel size of (4, 4, 4), a stride of (2, 2, 2) and a  padding of (1, 1, 1) while decoder convolutional layers have a kernel size of (3, 3, 3), a stride of (1, 1, 1) and a  padding of (1, 1, 1). A detailed schema of the VAE we use can be found in Figure~\ref{fig:vae}.

Our 3D convolutional VAE implementation relies on the open-source python library Pythae  \citep{chadebec2022pythae}. The model was trained on 200 epochs, with a learning rate of $10^{-5}$ using the ClinicaDL \citep{thibeau-sutreClinicaDLOpensourceDeep2022} software that aims to facilitate the use of the neuro-images with deep learning and improve reproducibility of the experiments.  
We trained the VAE on the Jean Zay high performance computer cluster with Nvidia Tesla V100 GPUs that have 32GB of memory, which allowed us to use a batch size of 8. It took approximately 10 hours to train each fold over all the epochs.

\section{Results}
\label{sec:results}

In this section, we present the results obtained when applying the proposed validation procedure to the pseudo-healthy reconstruction of 3D FDG PET images with a VAE for the detection of anomalies characteristic of Alzheimer's disease and other dementias. As a reminder, the validation procedure consists of four steps:
\begin{itemize}
    \item computing reconstruction metrics for images of healthy subjects to evaluate the quality of the reconstruction;
    \item measuring the healthiness of the reconstructed pseudo-healthy images using the simulation framework by comparing the pseudo-healthy reconstruction $\mathbf{\widehat{x'}}$ to the original scan from a healthy subject $\mathbf{x}$, and also by computing the newly introduced healthiness metric $H$;
    \item detecting regions of the brain containing anomalies using an atlas and validating this approach with the simulation framework;
    \item detecting anomalies on a real dataset of patients diagnosed with AD.
\end{itemize}

\subsection{Pseudo-healthy reconstruction performance for cognitively normal subjects}

The variational autoencoder is trained on six folds in order to evaluate the variance due to data splitting \citep{bouthillier2021accounting}. The results on the test set are summarized in Table~\ref{tab:result_cn}. We can first observe that the MSE is almost identical on the six folds with a mean-squared reconstruction error around $1.82 \times10^{-3}$. The performance measured with PSNR are also very similar, which is consistent as the PSNR is a function of the MSE. We also observed that the SSIM varies from 0.874 on average to 0.879 between the folds 4 and 2. 

\begin{table}[!htb]
    \caption{
        Reconstruction metrics obtained on the test set for images from healthy subjects over the 6 folds.
    }
    \begin{center}
    \renewcommand{\arraystretch}{1.25}
    \begin{tabular}{l|c|cccc}
    Dataset                         & Fold  & MSE ($\times10^{-3}$) \textdownarrow	& PSNR \textuparrow	& SSIM \textuparrow	& MS-SSIM \textuparrow	\\ \hline 
    \multirow{6}{*}{CN test set}    & 0     & $1.834\pm0.654$	& $27.527\pm1.080$	& $0.875\pm0.027$	& $0.944\pm0.015$		\\
                                    & 1     & $1.815\pm0.649$	& $27.572\pm1.074$	& $0.878\pm0.026$	& $0.944\pm0.014$		\\
                                    & 2     & $1.841\pm0.746$	& $27.532\pm1.113$	& $0.879\pm0.025$	& $0.944\pm0.015$		\\
                                    & 3     & $1.826\pm0.665$	& $27.547\pm1.079$	& $0.876\pm0.027$	& $0.943\pm0.014$		\\
                                    & 4     & $1.858\pm0.619$	& $27.456\pm1.031$	& $0.874\pm0.026$	& $0.944\pm0.014$		\\
                                    & 5     & $1.836\pm0.727$	& $27.546\pm1.134$	& $0.876\pm0.028$	& $0.943\pm0.015$		\\
    \end{tabular}
    \end{center}
\label{tab:result_cn}
\end{table}

For the following experiments, as it is difficult to interpret the results across several folds for a reconstruction task, we select a fold according to the SSIM, which is a perceptual metric substantially different from the loss being minimized. We show the results obtained with the fold 1, which presents an average SSIM on the validation set similar to that of the other folds, but has the highest minimum SSIM (see Appendix~\ref{sec:val}).

The metrics appear consistent with the reconstructions that are displayed in Appendix~\ref{sup:supp_recon_cn}. The reconstructions are acceptable for 3D full resolution images: the shape of the brain and the main structures, such as the ventricles, are well captured. However, smaller anatomical features such as cortical folds are not well reconstructed. 

\subsection{Evaluation of the model using the simulation framework}

\subsubsection{Results on simulated AD-like FDG PET images}
\label{sec:sim_results}

\begin{figure}[!htb]
    \centering
    \includegraphics[width=0.9\textwidth]{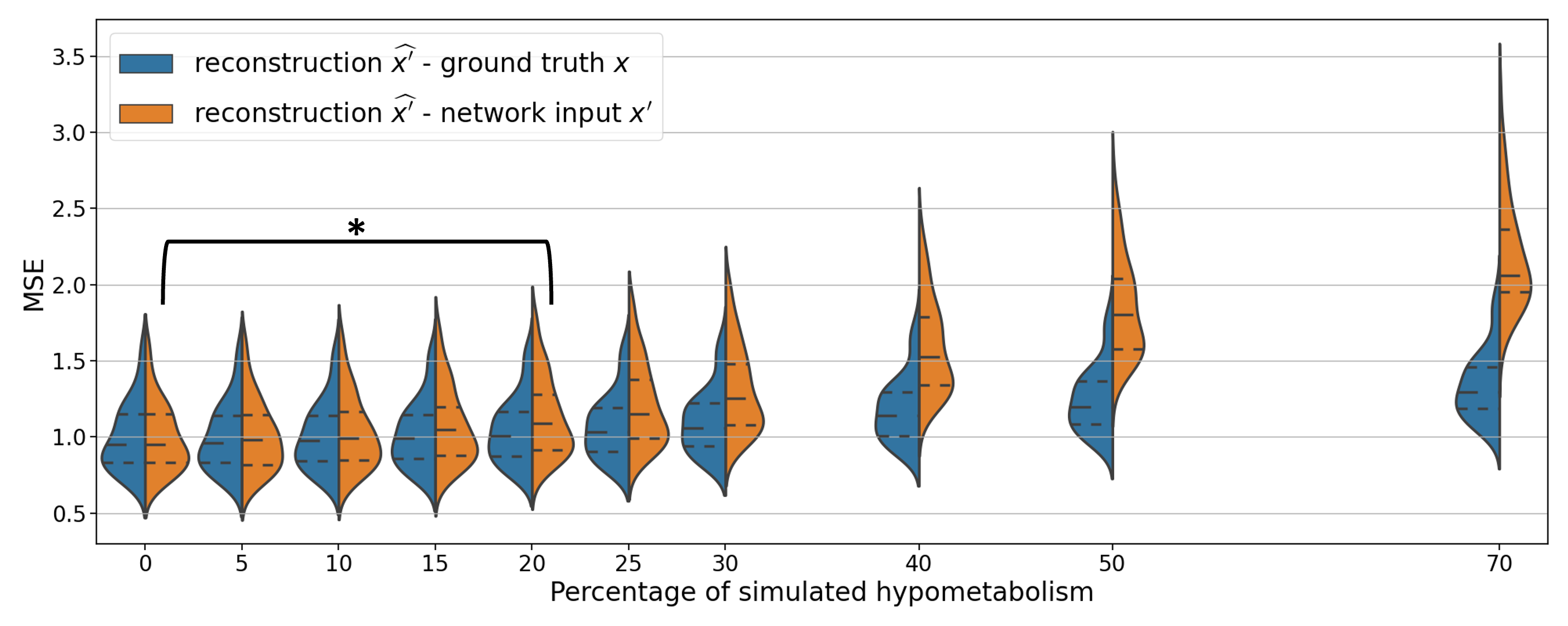}
    \caption{
    Evolution of the MSE with increasing degrees of hypometabolism simulating AD-like anomalies. We plot the distribution of the MSE between the pseudo-healthy reconstruction and the original image $MSE(\mathbf{x},\mathbf{\widehat{x'}})$ blue, and the MSE between the pseudo-healthy reconstruction and the simulated data $MSE(\mathbf{x'},\mathbf{\widehat{x'}})$ orange. Each MSE is normalized by the average MSE obtained when reconstructing from the original healthy images.
    }
    \label{fig:violin}
\end{figure}

To evaluate the impact of the anomaly severity on the ability of the VAE to reconstruct pseudo-healthy images, we simulated different degrees of hypometabolism from 5\% to 70\% (Figure~\ref{fig:violin}). We first remark that the reconstruction is satisfying until around 20\% of simulated hypometabolism as the MSE between the reconstruction $\mathbf{\widehat{x'}}$ and the ground truth $\mathbf{x}$ is almost constant.

Figure~\ref{fig:violin} also shows that the MSE between the simulated input image $\mathbf{x'}$ and the output $\mathbf{\widehat{x'}}$ is higher for more severe anomalies. This confirms that the model cannot reconstruct well highly abnormal areas. 

We then compared the reconstruction error that was obtained for the simulated data (i.e., $MSE(\mathbf{x'},\mathbf{\widehat{x'}})$, in orange in Figure~\ref{fig:violin}) with the error that exists between the real healthy images from the CN test set and the reconstructions obtained from the simulated data  (i.e., $MSE(\mathbf{x},\mathbf{\widehat{x'}})$, in blue in Figure~\ref{fig:violin}).  We remark that $MSE(\mathbf{x'},\mathbf{\widehat{x'}})$ does not increase as much as $MSE(\mathbf{x},\mathbf{\widehat{x'}})$ with the hypometabolism severity. This means that the reconstruction $\mathbf{\widehat{x'}}$ is more similar to the ground truth $\mathbf{x}$ than the abnormal simulated image $\mathbf{x'}$. However, the error still increases between the reconstruction $\mathbf{\widehat{x'}}$ and the ground truth $\mathbf{x}$, meaning that a healthy image cannot be totally recovered when the anomalies are too intense. We also observe that for low degree hypometabolism (\textless 20\%), both MSEs are similar. This means that the residual error due to the model imperfect reconstruction dissimulates the reconstruction error due to low degree anomalies. To confirm our observations, we computed a t-test assessing whether there was a significant difference in MSE between the reconstruction from the simulated input $\mathbf{\widehat{x'}}$ and the ground truth $\mathbf{x}$ using images with various degrees of anomalies. The p-values were corrected for multiple comparisons using the Bonferroni method with 10 comparisons. The difference in MSE becomes significant (p-value\textless 0.005) for anomalies of degree 20\% and above. This shows that we can detect hypometabolism around 25\% using the residual error, which corresponds to the average difference in metabolism between CN subjects and AD patients in a region of interest relevant to AD in ADNI \citep{Landau2012AmyloidDeposition}, knowing that this dataset includes patients at a very early stage of the disease.

\begin{figure}[!htb]
    \centering
    \includegraphics[width=\textwidth]{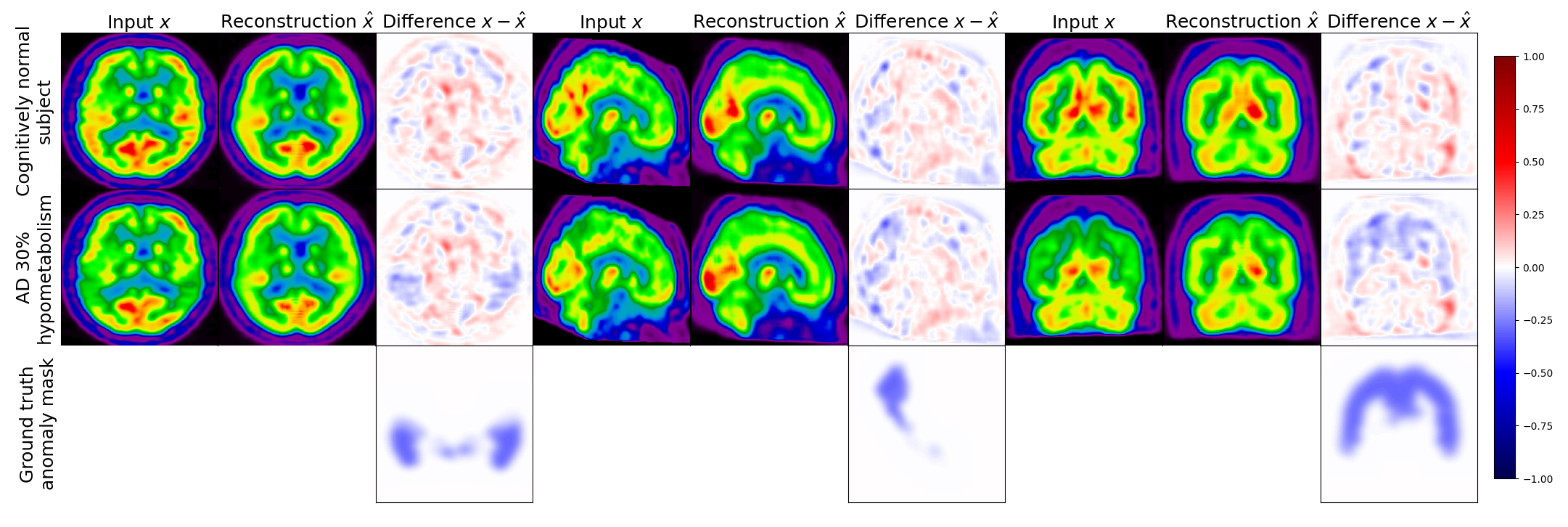}
    \caption{
        Example of results obtained from a real image  of a CN subject (top row) and an image simulating AD hypometabolism based on the same CN subject (bottom row). For each plane, the first image is the input, the second one the model's reconstruction and the third one the difference (input - reconstruction). 
    }
    \label{fig:results}
\end{figure}

Figure~\ref{fig:results} displays the real image of a CN subject $\mathbf{x}$ and its pseudo-healthy reconstruction $\mathbf{\hat{x}}$, as well as the simulated AD version $\mathbf{x'}$ (with a hypometabolism degree of 30\%) and its reconstruction $\mathbf{\widehat{x'}}$ obtained from the same CN subject, together with the residual images. We observe that the input and output images of the CN subject are quite similar, both the shape of the brain and the uptake distribution look alike. The differences are due to the model imperfect reconstruction and correspond to the minimal error that it can achieve.

When feeding the simulated hypometabolic image $\mathbf{x'}$ to the model, we observe that the reconstructed image $\mathbf{\widehat{x'}}$ looks healthier than the input image. The areas highlighted in blue in the residual map correspond to the regions where hypometabolism was simulated.

Another interesting point is that both images reconstructed from the same CN subject $\mathbf{\hat{x}}$ and $\mathbf{\widehat{x'}}$ are almost identical with a SSIM of 0.987. This shows that the model reconstructs almost the same image for the same subject whether the input image is healthy or presents anomalies.

\subsubsection{Results when simulating various types of dementia}

In this section, the degree of hypometabolism is set to 30\% but the brain region where it is simulated changes to reflect various types of dementia. We report in Table \ref{tab:simulateddiseases} the different reconstruction metrics computed between the original images from CN subjects in the test set $\mathbf{x}$ and the images reconstructed from the hypometabolic scans simulating the different types of dementia $\mathbf{\widehat{x'}}$. We observe that the metrics are similar for all the simulated dementias, which means that the model can generalize to anomalies with different locations and shapes, as well as different severity degrees as we showed previously.

\begin{table}[!htb]
    \caption{
        Reconstruction metrics computed between the original healthy PET scans from CN subjects in the test set and the images reconstructed with the 3D VAE from the hypometabolic scans simulating different types of dementia.
    }
    \label{tab:simulateddiseases}
    \begin{center}
    \renewcommand{\arraystretch}{1.25}
    \begin{tabular}{l|c c c c}
        % \rule[-1ex]{0pt}{3.5ex}   
        Simulated dementia  & MSE ($\times10^{-3}$) \textdownarrow & PSNR \textuparrow & SSIM \textuparrow & MS-SSIM \textuparrow \\ \hline
        AD	& $2.230\pm0.655$	& $26.646\pm0.996$	& $0.848\pm0.050$	& $0.938\pm0.015$	\\
        bvFTD	& $2.268\pm0.686$	& $26.584\pm1.042$	& $0.849\pm0.051$	& $0.940\pm0.015$	\\
        PCA	& $2.090\pm0.698$	& $26.962\pm1.119$	& $0.851\pm0.051$	& $0.942\pm0.015$	\\
        lvPPA	& $2.073\pm0.680$	& $26.992\pm1.104$	& $0.850\pm0.051$	& $0.941\pm0.015$	\\
        nfvPPA	& $2.093\pm0.692$	& $26.953\pm1.107$	& $0.851\pm0.051$	& $0.941\pm0.015$	\\
        svPPA	& $2.029\pm0.703$	& $27.101\pm1.150$	& $0.852\pm0.051$	& $0.942\pm0.016$	\\
    \end{tabular}
    \end{center}
\end{table}

We also computed the metrics between the images reconstructed from the original healthy scans $\mathbf{\hat{x}}$ and the images reconstructed from the simulated hypometabolic scans $\mathbf{\widehat{x'}}$ in Table~\ref{tab:ssimgtrecon}. Both reconstructions are almost identical with an SSIM on average superior to 0.99. We can conclude from this experiment that the model is able to reconstruct the healthy version of an image independently of the nature of the dementia that causes the anomaly.

\begin{table}[!htb]
    \begin{center}
    \caption{
        Structural similarity between the pseudo-healthy reconstruction $\mathbf{\widehat{x'}}$ and the reconstruction from the healthy image $\mathbf{\hat{x}}$ for the different dementias simulated.
    }
    \label{tab:ssimgtrecon}
    \renewcommand{\arraystretch}{1.25}
    \begin{tabular}{l|c}
        Simulated dementia  &  SSIM \textuparrow \\ \hline
        AD	    & $0.9878\pm0.0014$  \\
        bvFTD	& $0.9921\pm0.0013$	\\
        PCA	    & $0.9974\pm0.0003$	\\
        lvPPA	& $0.9937\pm0.0008$	\\
        nfvPPA	& $0.9964\pm0.0005$	\\
        svPPA	& $0.9995\pm0.0002$	\\
    \end{tabular}
    \end{center}
\end{table}

\subsubsection{Measuring healthiness of a pseudo-healthy reconstruction}

We computed the proposed healthiness metric for the different simulation experiments: on the test sets simulating AD with various intensity degrees and on the test sets simulating the different dementia subtypes. 

\begin{figure}[!htb]
    \centering
    \includegraphics[width=0.9\textwidth]{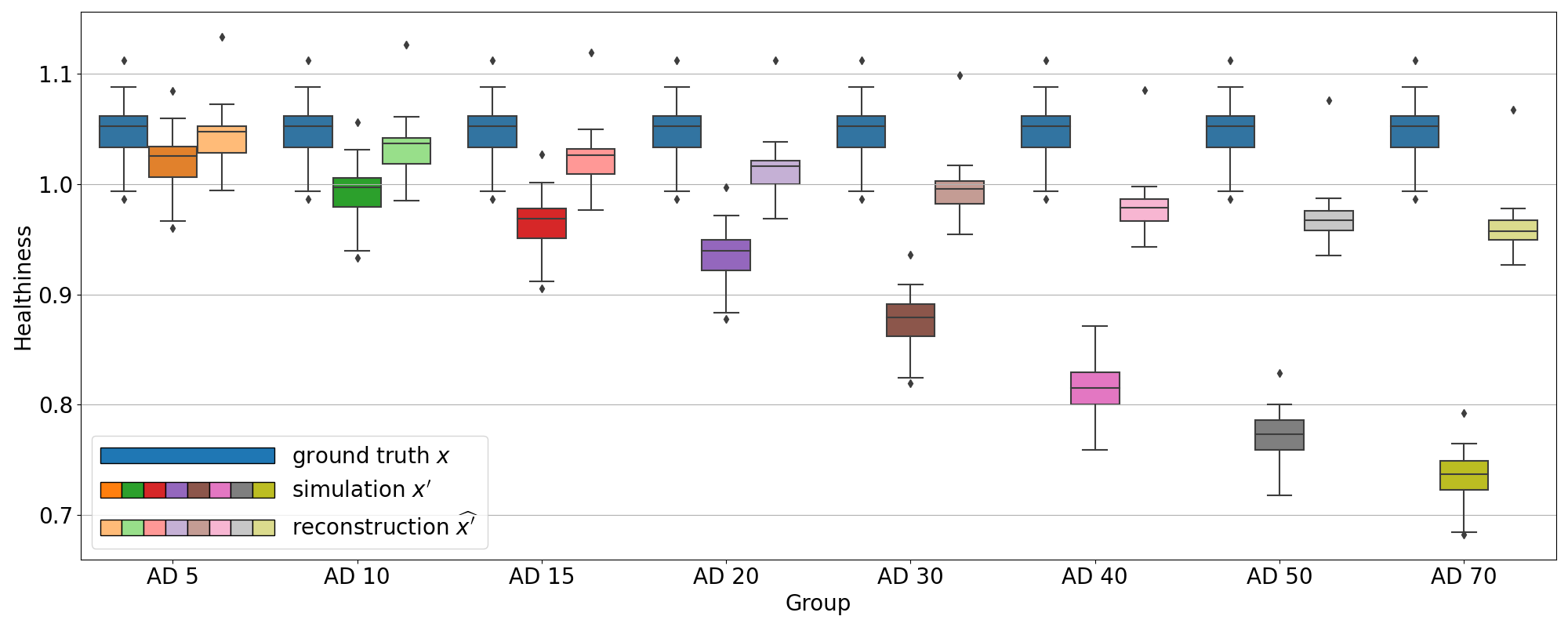}
    \caption{
    Evolution of the distribution of the healthiness metric computed for the ground truth healthy images, their corresponding simulated images and their pseudo-healthy reconstructions when increasing the percentage of AD-like simulated hypometabolism.
    }
    \label{fig:healthiness}
\end{figure}

We can see in Figure~\ref{fig:healthiness} that the healthiness score for the original PET scans from CN subjects $\mathbf{x}$ ranges between 0.99 and 1.08, which can define a baseline of what we can consider as healthy with this metric. As expected, we observe that the healthiness of simulated images $\mathbf{x'}$ is lower than that of the original image $\mathbf{x}$. At 5\% of simulated hypometabolism, the score is still between 0.97 and 1.06, so it can be considered as healthy, which is coherent for very low anomaly severity. From 15\% of simulated hypometabolism, we can clearly see that the healthiness score drops and become much lower than that of the healthy images: it is inferior to 1.0 for 15\% of simulated hypometabolism and it is between 0.82 and 0.91 for 30\% of simulated hypometabolism. The important point is that the healthiness score of the reconstruction $\mathbf{\widehat{x'}}$ is always superior to the one of the simulated image $\mathbf{x'}$. We can see that it is even really close to the healthiness of the original image $\mathbf{x}$: for 30\% of simulated hypometabolism the healthiness of reconstructed images is between 0.95 and 1.03. 

\begin{figure}[!htb]
    \centering
    \includegraphics[width=0.9\textwidth]{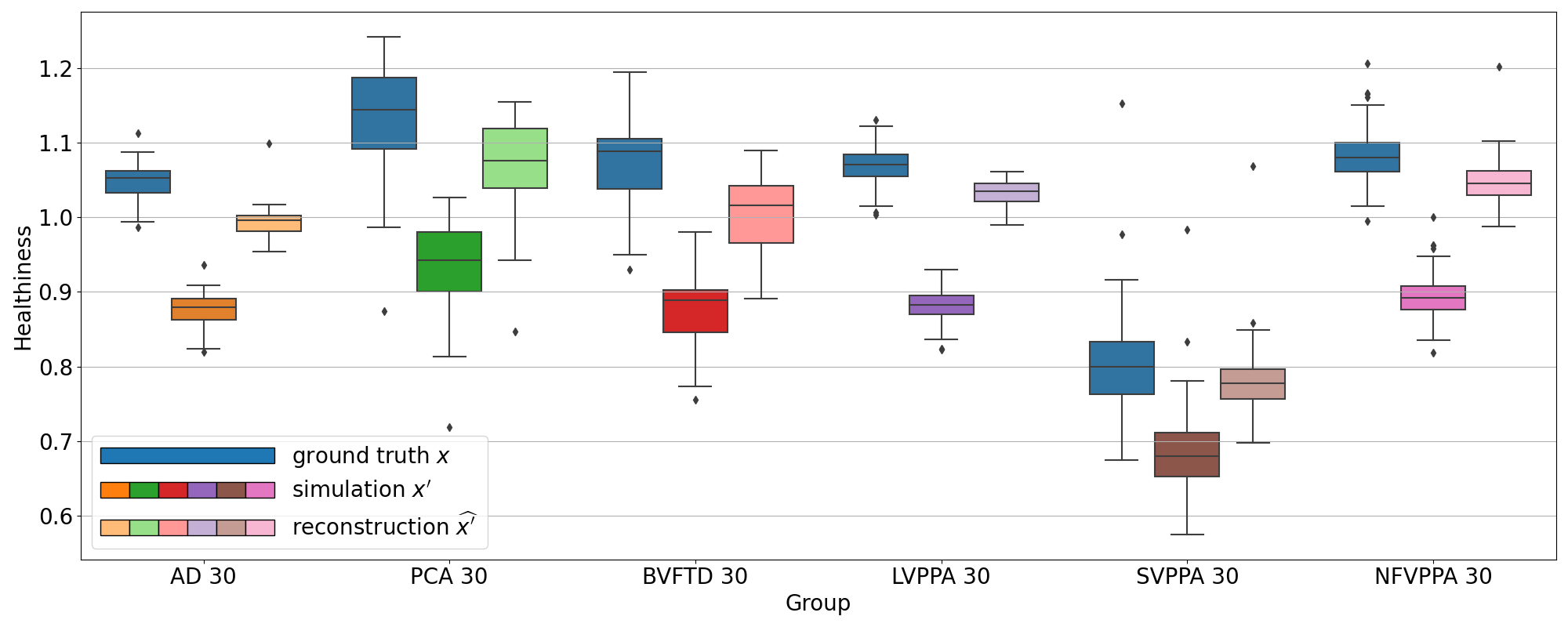}
    \caption{
    Evolution of the distribution of the  healthiness metric computed for the ground truth healthy images, their corresponding the simulated images and their pseudo-healthy reconstructions for different dementias simulated at 30\%.
    }
    \label{fig:healthiness_path}
\end{figure}

We observe the same behavior for all the other simulated dementias in Figure~\ref{fig:healthiness_path}. However, we note that the healthiness of the ground truth, i.e. that obtained for images of CN subjects, varies depending on the dementia simulated because the mask used to compute it differs. For example, in the case of svPPA, the healthiness of the ground truth is lower than that of AD (between 0.67 and 0.92). This can be explained by the fact that the mask used for svPPA is located in the temporal pole, where FDG uptake is lower compared to other regions, even on healthy images, as we can see in Figure~\ref{fig:anomaly_regions}. However, we can still observe that the healthiness is lower on the simulated image compared to the healthy image and almost equal on the reconstruction.

This method relies on the simulation framework and can only be used to evaluate the model performance. However we would like a method or metric that allows clinicians to know if an image presents anomalies, and possibly localize them. 

\subsubsection{Anomaly detection applied to simulated data}

To detect anomalies in real images of patients, i.e., without having to rely on the simulation framework, we proposed to compute the mean uptake in regions of an atlas and compared the values between input and output. If the value in the reconstructed image $\mathbf{\hat{x}}$ is close to the input image $\mathbf{x}$, then the region is not likely to be abnormal, otherwise, if the regional uptake in the reconstructed image $\mathbf{\hat{x}}$ is significantly different from the one in the original image $\mathbf{x}$, then there may be an anomaly. We validated this assumption using the simulation framework.

\begin{figure}[!htb]
    \centering
    \includegraphics[width=\textwidth]{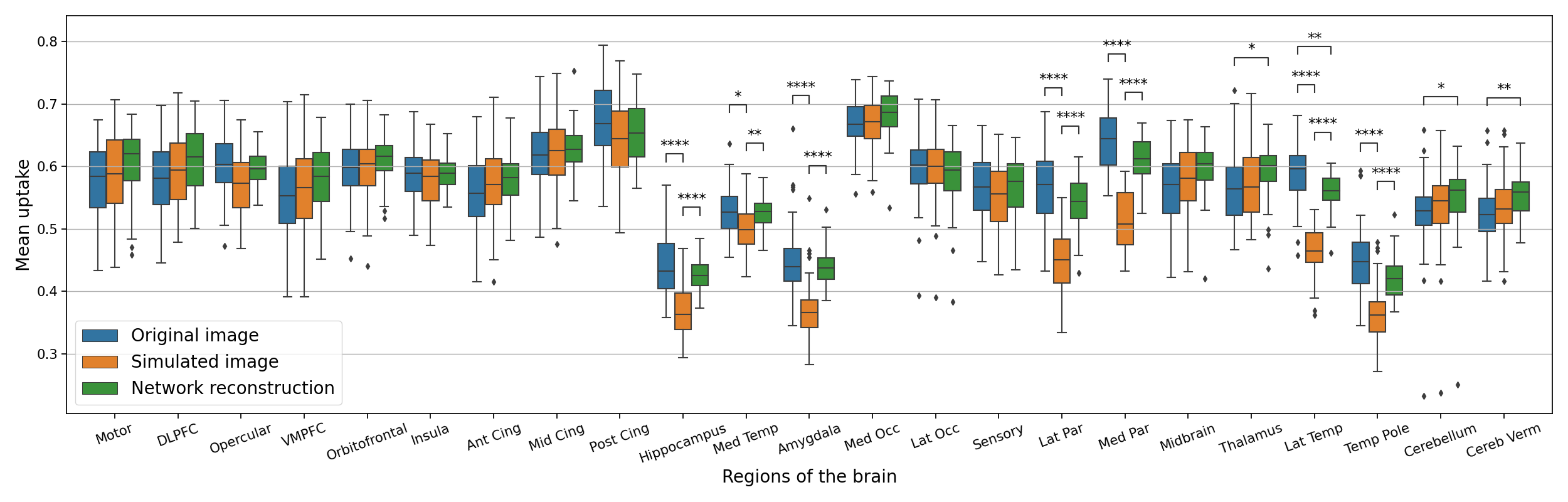}
    \caption{
    Distribution of the mean FDG PET uptake in different regions of the brain: comparison between the CN subjects from the test set, their AD-like hypometabolic simulation, and their pseudo-healthy reconstruction.
    }
    \label{fig:anomaly_regions}
\end{figure}

In Figure~\ref{fig:anomaly_regions}, we plotted the average uptake in the regions of the atlas we used and compared these values between the original image $\mathbf{x}$, the simulated one $\mathbf{x'}$ and the reconstruction $\mathbf{\widehat{x'}}$ using a Wilcoxon-Mann-Whitney test corrected with Bonferroni for multiple comparisons using the statannotations package~(\url{https://statannotations.readthedocs.io/en/latest/index.html}). First, we remark that the average uptake is not consistent between all the regions of the brain, so we cannot really compute a shift from an average value for the whole brain, but we have to analyze the average uptake for every region. We can then observe that the average uptake is not significantly different between the original image $\mathbf{x}$ and the simulated one $\mathbf{x'}$, the original image $\mathbf{x}$ and the reconstruction $\mathbf{\widehat{x'}}$, and the simulation $\mathbf{x'}$ and its reconstruction $\mathbf{\widehat{x'}}$ for most of the regions, except the hippocampus, the amygdala, the parietal lobe and the temporal lobe. These regions correspond to the regions used to simulate anomalies corresponding to AD. We can see that the average uptake is lower on the simulated image $\mathbf{x'}$ compared to the original image $\mathbf{x}$. This is expected as the hypometabolism simulation consists in lowering the intensity in those regions. We can also see that the average uptake is significantly higher in these regions compared to the average uptake on the simulated images. Without a priori knowledge on the nature of the anomaly we want to detect, we can see that on this test set, it is likely to be abnormal in these regions. This corroborates with the regions we actually used to simulate hypometabolism.

Now that we extensively used the simulation framework to validate our model on different aspects: pseudo-healthy reconstruction, anomaly detection, generalization to anomalies of different intensities, locations and shapes, we will examine the results on the images of AD patients from the ADNI database.

\subsection{Results on AD patients from the ADNI dataset}

\begin{figure}[!htb]
    \centering
    \includegraphics[width=\textwidth]{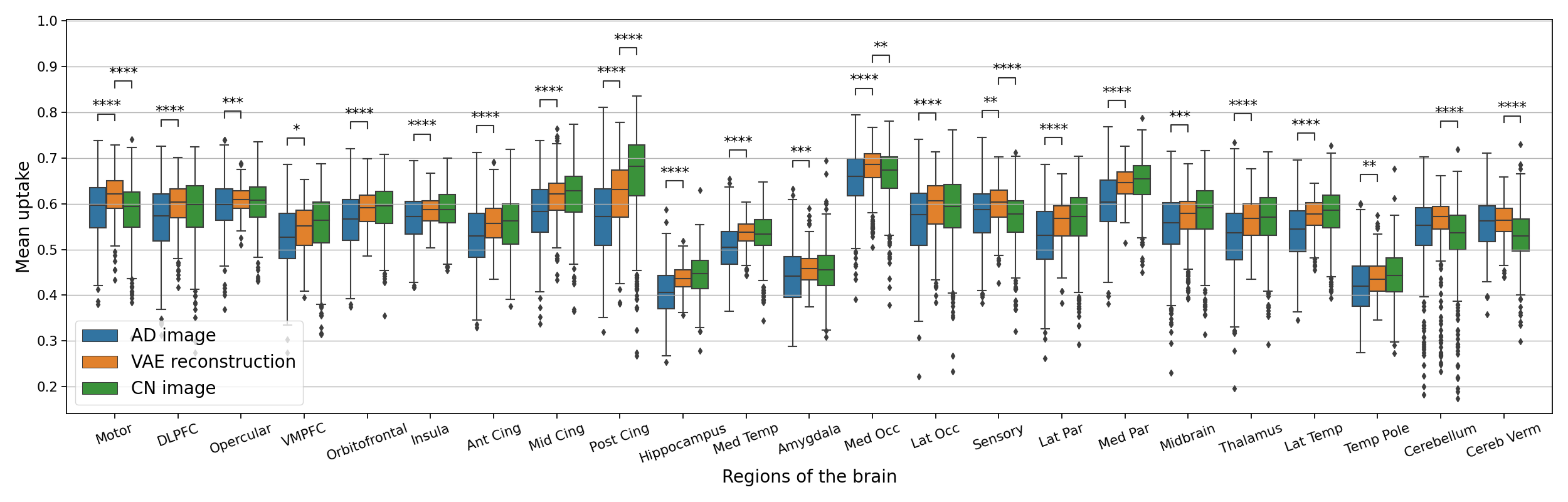}
    \caption{
        Distribution of the mean FDG PET uptake in different regions of the brain: comparison between the original image from AD patients, their pseudo-healthy reconstruction and the CN population.
    }
    \label{fig:AD_anomaly}
\end{figure}

The results of the anomaly detection method applied to real AD patients are reported in Figure~\ref{fig:AD_anomaly}. We can observe that in general the average uptake is higher in the pseudo-healthy reconstruction. We cannot really detect abnormal areas even though we can see that regions such as the posterior cingulate, hippocampus, parietal lobe, lateral temporal lobe seem to be regions with the largest differences between the AD patients and their pseudo-healthy reconstruction. This global analysis can help us describe the cohort and understand at the population level the shift from the CN population. However it is not really an image-level anomaly detection tool. For that, we need to observe individually each image. Some examples of reconstructions from AD patients are displayed in Appendix~\ref{sup:recon_ad}.

To be consistent with the literature, we also compared the reconstruction metrics computed on images of AD patients with the metrics computed on images of CN subjects. We report the results in Table \ref{tab:cnvsad}. We can see that the reconstruction performance is slightly better on the images from CN subjects, which is a positive point because, as the model should correct anomalies on images from AD patients, the reconstruction error is expected to be higher. However the difference is not large enough to separate healthy controls from AD patients.

\begin{table}[!htb]
    \caption{
        Comparison of reconstruction metrics computed on test set with healthy subjects and test set with patients with AD.
    }
    \label{tab:cnvsad}
    \begin{center}
    \renewcommand{\arraystretch}{1.25}
    \begin{tabular}{l|c c c c}
        Simulated dementia  & MSE ($\times10^{-3}$) \textdownarrow & PSNR \textuparrow & SSIM \textuparrow & MS-SSIM \textuparrow \\ \hline
        CN	& $1.815\pm0.649$	& $27.572\pm1.074$	& $0.878\pm0.026$	& $0.944\pm0.014$	\\
        AD	& $2.554\pm1.391$	& $26.272\pm1.560$	& $0.853\pm0.045$	& $0.928\pm0.025$	\\
    \end{tabular}
    \end{center}
\end{table}

\subsection{Latent space analyses}

Now that we extensively tested the model under various conditions, we would like to better understand its behavior and interpret the results. Indeed, for instance, we would like to understand why the reconstruction  $\mathbf{\widehat{x'}}$ obtained from a simulated abnormal image $\mathbf{x'}$ looks similar to the reconstruction  $\mathbf{\hat{x}}$ obtained from the original input image $\mathbf{x}$. Actually, there is no reason that would a priori explain why the reconstruction of an abnormal image would be realistic and correspond to a healthier version of the input image. We could imagine that, like for out-of-distribution detection methods, the model would not reconstruct the input image at all. This is why will use the latent space representation to study our model and understand what the VAE learns. In the latent space, all the input images are projected into a one dimension vector space of size 256 through the encoder. The advantage of the VAE is that the latent space is consistent, that is to say that, in theory, the latent representation of the images are organized with respect to the image distribution. We will verify whether this is actually the case.

\begin{figure}
    \centering
    \begin{subfigure}[b]{0.49\textwidth}
        \centering
        \includegraphics[width=\textwidth]{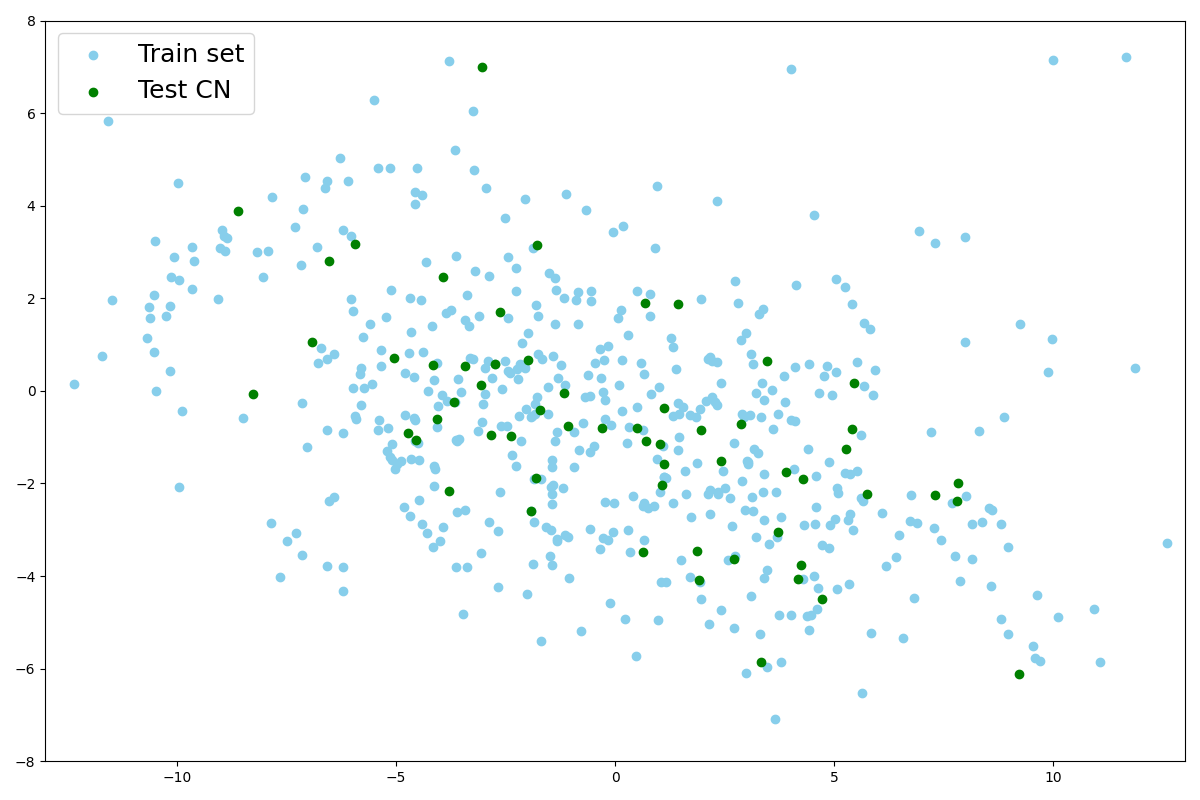}
        \caption{}    
        \label{fig:emb_train}
    \end{subfigure}
    \hfill
    \begin{subfigure}[b]{0.49\textwidth}  
        \centering 
        \includegraphics[width=\textwidth]{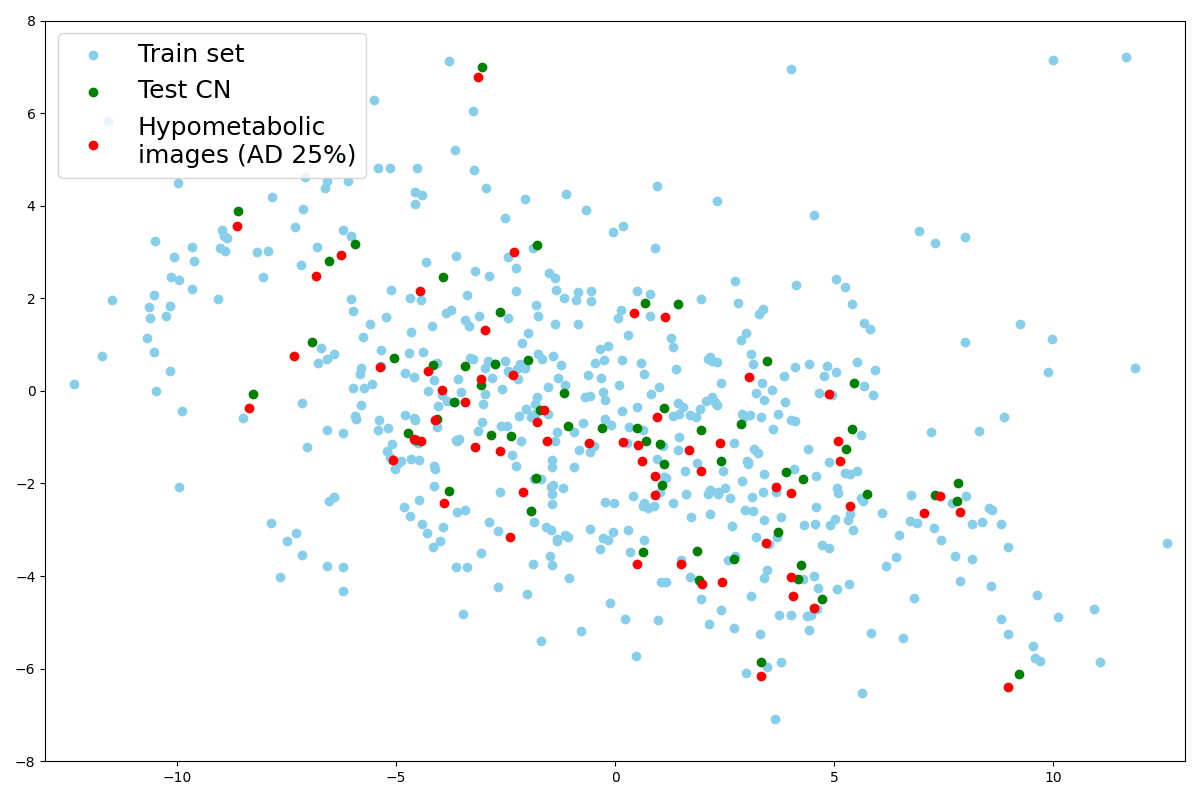}
        \caption{} 
        \label{fig:emb_CN}
    \end{subfigure}
    \vskip\baselineskip
    \begin{subfigure}[b]{0.49\textwidth}   
        \centering 
        \includegraphics[width=\textwidth]{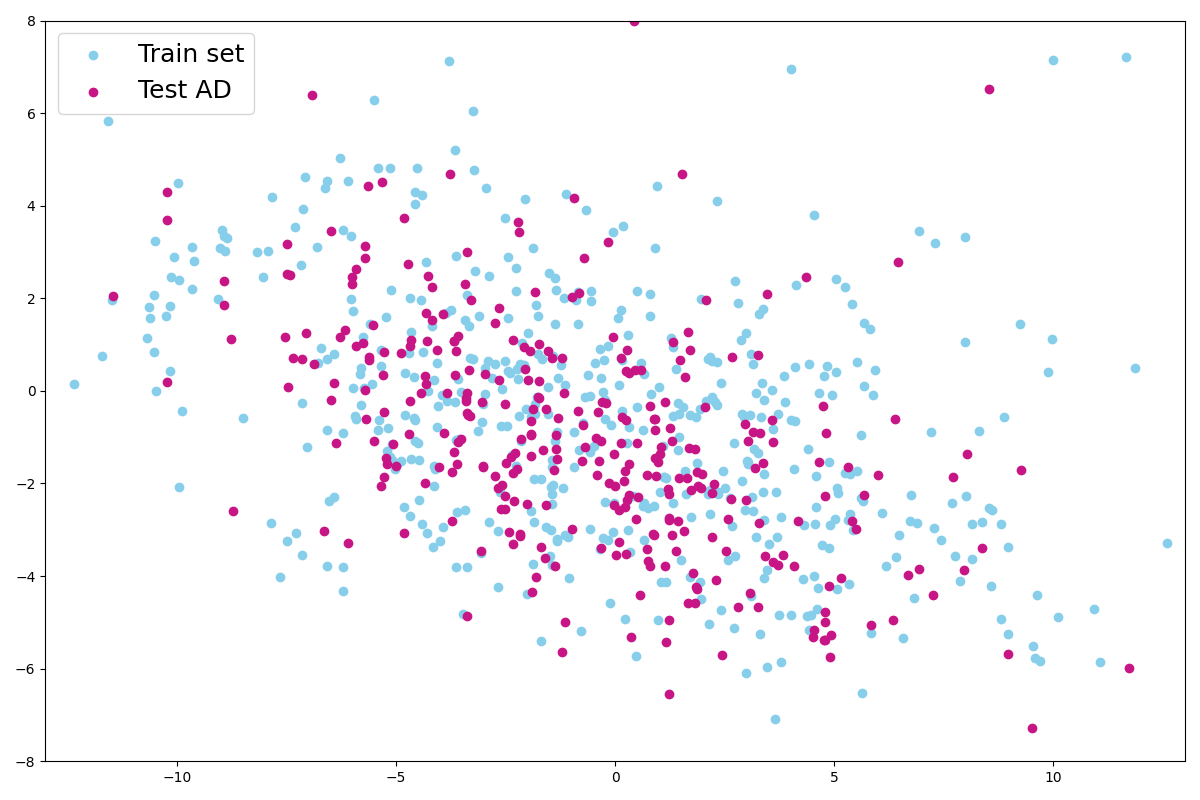}
        \caption{}
        \label{fig:emb_AD}
    \end{subfigure}
    \hfill
    \begin{subfigure}[b]{0.49\textwidth}   
        \centering 
        \includegraphics[width=\textwidth]{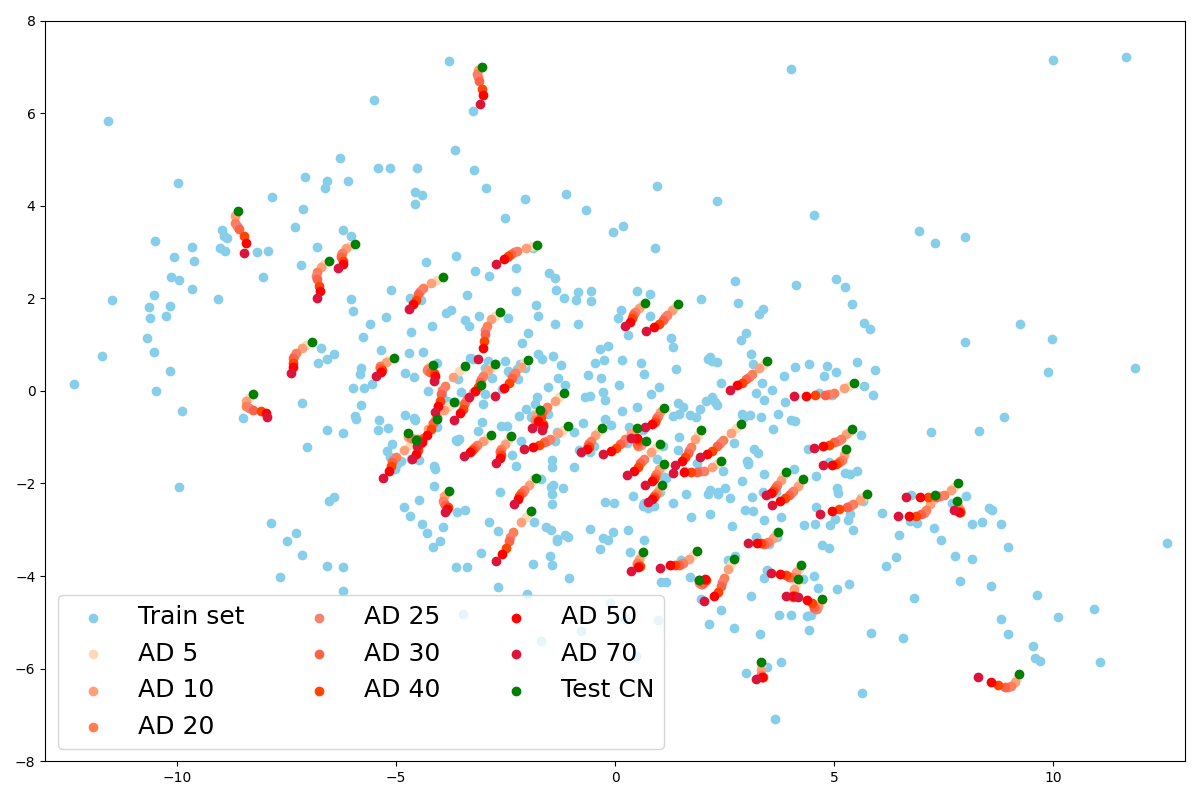}
        \caption{}
        \label{fig:emb_AD_pr}
    \end{subfigure}
    \caption{
        Latent space representation (first two PCA components). The latent distribution of the train set learned by the VAE (in blue) is compared to the latent distribution inferred on the test set with CN subjects (a), test set with AD patients (c), and image simulating AD-like hypometabolism with progression from 5\% to 70\% (b, d).
    } 
    \label{fig:latent_pca}
\end{figure}

We first used a PCA to reduce the dimension of the latent space and extract the principal components so we can observe the latent distribution into a 2D plot. We fitted the PCA on the train set as we can see in Figure~\ref{fig:emb_train}.

We then predict the principal components of the latent representation of images from the CN test set with the same PCA, as well as their hypometabolic version simulating AD. A remarkable point is that the projection is almost the same for images that have been simulated from this test set as shown with the paired points in Figure~\ref{fig:emb_CN}. This explains why their reconstruction are almost identical as we noticed in section \ref{sec:sim_results}. Indeed, the decoder will reconstruct two similar images from two similar latent vectors. 

We also project latent vectors of images from the AD test set, and we can see that the points are in the same area of the latent space (Figure~\ref{fig:emb_train}). This validates our hypothesis that images presenting anomalies (real or simulated) are projected into the healthy images' latent distribution that was learned on the training set. We can observe that in practice $q_{\Phi}(\mathbf{z \mid x_h}) \approx q_{\Phi}(\mathbf{z \mid x_p})$ and that the latent representation $\mathbf{z}$ is a small sphere in the latent space. 

Another interesting point is that the latent space seems to capture the simulated progression of AD. We observe in Figure~\ref{fig:emb_AD_pr} that the principal component vectors of AD simulated images are aligned in the latent space, near the original image latent representation and ordered by severity.

\begin{figure}[!htb]
    \centering
    \includegraphics[width=0.9\textwidth]{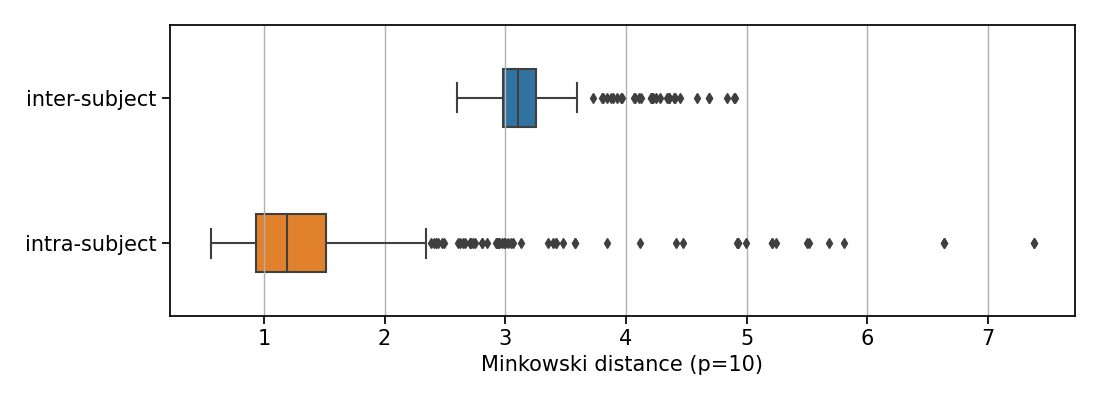}
    \caption{
        Boxplot showing the distribution of the Minkowski distance computed between the latent representation of images from the same subject (intra-subject) and between the closest latent representation of images from other subjects (inter-subject).
    }
    \label{fig:intervsintra}
\end{figure}

We then used the Minkowski distance to compute intra-subject and inter-subject distances in the latent space. We first observe that, for a given image of a subject, all the closest images in the latent space are images of the same subject, acquired during other visits. Figure~\ref{fig:intervsintra} displays the box plots of the mean distance between the latent representation of an image and that of the other images of the same subject, and the mean distance between an image and the five closest latent representations of images that are not from the same subject. The difference between intra-subject and inter-subject distances is statistically significant (p-value $\ll 0.005$ according to a Mann-Whitney U test). This clearly indicates that all the images from a same participant are very close in the latent space compared to the average distance between two images from different participants.

\begin{figure}
     \centering
     \begin{subfigure}[b]{0.49\textwidth}
         \centering
         \includegraphics[width=\textwidth]{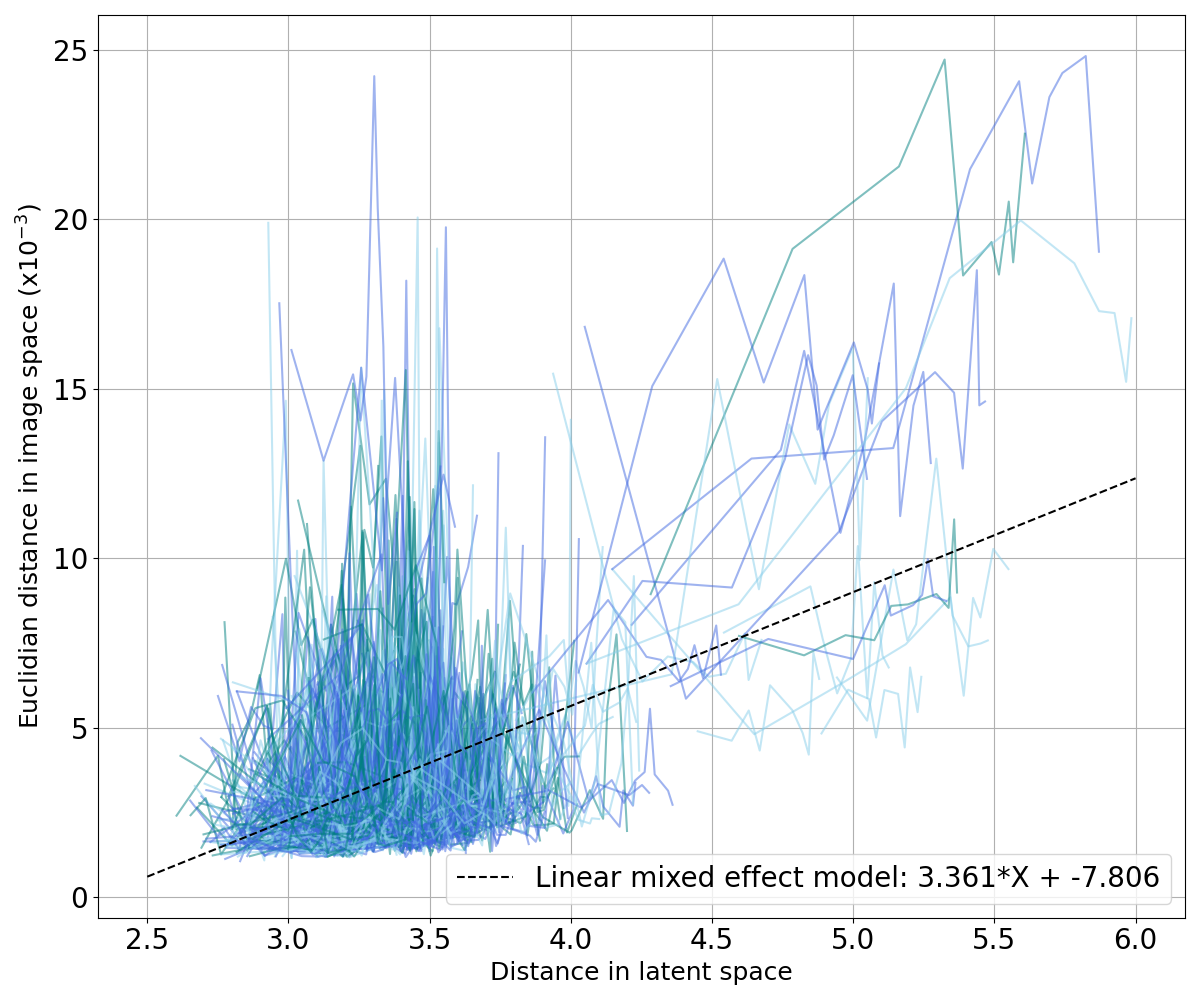}
         \caption{Distance in latent space (Minkowski) compared to distance in image space (MSE)}
         \label{fig:msevslatent}
     \end{subfigure}
     \hfill
     \begin{subfigure}[b]{0.49\textwidth}
         \centering
         \includegraphics[width=\textwidth]{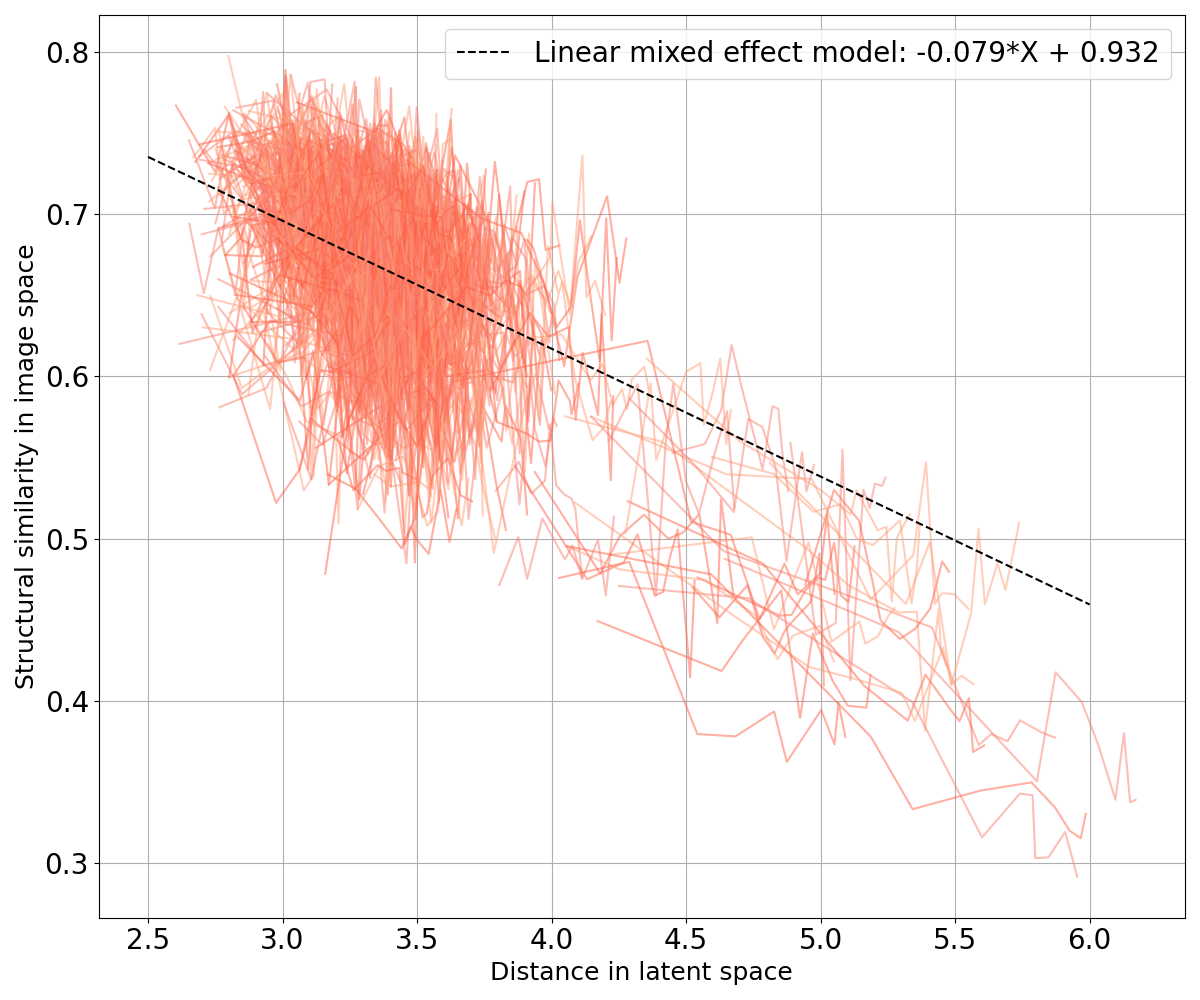}
         \caption{Distance in latent space (Minkowski) compared to similarity in image space (SSIM)}
         \label{fig:ssimvslatent}
     \end{subfigure}
    \caption{
        Evolution of the MSE and the SSIM compared with the Minkowski distance in the latent space. Each curve represents an image $i$ and comprises ten points. Each point of the curve corresponds to the distance $D_{ij}$ of this image $i$ with the $j^{th}$ closest images in the latent space, j being in \{1, 6, 11, 16...41, 46\}. Detailed results of linear mixed effect models are in Appendix~\ref{sec:lmm_results}.
    }
    \label{fig:imgvslatent}
\end{figure}

We also want to ensure that, if two points are close in the latent space, their corresponding images are close in the image space; and similarly, if they are far in the latent space, their corresponding images are not similar. We plot curves representing the evolution of the MSE and the SSIM with regard to their Minkowski distance in the latent space (Figure~\ref{fig:imgvslatent}) and fit two linear mixed effects models on both sets of curves to observe the general tendency: one for the MSE in Figure~\ref{fig:msevslatent} and one for the SSIM in Figure~\ref{fig:ssimvslatent}. We can see that the distance in the latent space increases when the MSE between two images grows and when the SSIM between the two images decreases. In other words, similar images have close latent representations.

\section{Discussion}

In this study, we proposed an in-depth validation procedure for pseudo-healthy synthesis with deep generative models in the context of UAD. This evaluation method relies on a simulation framework and is suited for applications for which ground truths are not available to measure the performance of the model. This procedure helps to extensively test a model on different aspects: the quality of the reconstruction, the healthiness of the reconstructed images, and the possibility to detect anomalies on both simulated and real data. We applied this procedure to evaluate the ability of a 3D VAE to detect anomalies related to Alzheimer's disease on 3D brain FDG PET data from the ADNI database.

As a first evaluation step using images from healthy subjects and common reconstruction metrics, we confirmed that the VAE was able to reconstruct subject-specific 3D FDG PET. However, the quality of the reconstruction can be improved as we observe that the VAE does not generate the sharp details of the image very well. The reconstruction is quite blurry, which is expected as the model is rather simple and generating high resolution 3D images is a difficult task. The results are still satisfactory for FDG PET images at this resolution since they are smooth by nature, but we can imagine that the VAE will be limited on other modalities that have a lot of structural details such as anatomical MRI. Even though the reconstruction quality may be sufficient to detect anomalies, it would be difficult to qualify the reconstruction as being pseudo-healthy at this stage of the evaluation \citep{baur_autoencoders_2021}. Moreover, using only reconstruction error to differentiate between healthy subjects and patients is not robust enough. The major weakness is that little intense anomalies might be difficult to detect for several reasons: first the model is able to reconstruct a meaningful image even though the image is abnormal, the reconstruction error due to abnormal regions will be drowned in the reconstruction noise due to model imperfection, and the reconstruction metrics being computed on the whole image may not highlight significant difference \citep{meissen2021pitfalls}. Nevertheless, in the case of AD, the typical variation of metabolism in relevant regions is around 25\% \citep{landau_amyloid_2014}, which corresponds to the intensity degree of anomalies that we can detect using the reconstruction error (Figure~\ref{fig:violin}).

To overcome the absence of ground truth anomaly masks for the evaluation of the model, we introduced a framework to simulate different kinds of dementias from images of healthy subjects. Having such pairs of diseased and healthy images allowed us to measure the reconstruction error between the pseudo-healthy reconstruction and the original image from the healthy subject. We showed that the pseudo-healthy reconstruction is more similar to the original image from the healthy subject than from the hypometabolic simulated image (Figure~\ref{fig:violin}).

To push the evaluation further, we also showed that the reconstructed images are looking healthy by introducing a new healthiness metric, which we validated thanks to the simulation framework. This analysis showed that the relative average uptake in the region used to simulate hypometabolism compared to the other regions of the brain is higher on the reconstruction, which means that the VAE can reconstruct a pseudo-healthy image (Figure~\ref{fig:healthiness} and Figure~\ref{fig:healthiness_path}). Actually, if the simulated hypometabolism is reasonable (\textless 30\%), the healthiness of the reconstructed pseudo-healthy images is similar to that obtained for the original image from a healthy subject. We also simulated dementias other than AD and showed that the VAE was indeed able to generalize to anomalies in different parts of the brain. This is an important point as many diseases are rare, so it is impossible to detect them using a traditional supervised machine learning approach due to lack of data.

We do not only rely on the simulated data to estimate the performance of the model, but we also use the image from AD patients to test the model in a more realistic context. Using the anomaly metric, we see that the pseudo-healthy reconstructions of images from real AD patients seem to have an average uptake similar to the healthy population (Figure~\ref{fig:AD_anomaly}. Unfortunately, at this stage, the individual analysis remains only visual and not quantitative since there is no ground truth healthy image for these patients, nor lesion masks. 

In most applications of pseudo-healthy synthesis for UAD, the performance of the model on real diseased images is measured with a similarity metric using a ground truth anomaly mask. This has the advantage of giving a quantitative measure for anomaly detection, however this does not really measure if the reconstructed image is pseudo-healthy. Evaluating models on anomaly detection only may lead to incomplete and biased evaluation as the model might not be able to detect every kind of anomalies. In our case, the final validation step would be to ask a clinician to evaluate the healthiness of the reconstructed images. However, having a clinician manually rate images is very time consuming and can be expensive, thus the simulation framework is an interesting solution as a first validation step to ensure that the model evaluated is worth asking expert advice.

The main advantage of using the proposed simulation framework is the possibility of quantitatively measuring the performance of the model using metrics (reconstruction and healthiness). This is crucial for further evaluation when one needs to compare models. To illustrate this, we trained a Unet model with a similar architecture to the one of the VAE. The results are reported in Appendix~\ref{sec:unet}. They highlight the importance of not only relying on reconstruction metrics and observations and demonstrate that simulated data can be useful to identify models that are not suited for pseudo-healthy reconstruction.

It also allows testing the model in many different conditions, with various kind of anomalies, and validate the fact that the model can generalize well. Another benefit is that this may lead to more robust anomaly detection by the clinician as it may be difficult to be vigilant on the whole image when manually inspecting a 3D scan.

One weakness of using simulated data is that they might not be very realistic. However we can clearly see that the simulated images are abnormal, which still allows evaluating performance, even though they are not totally realistic. For an even more comprehensive assessment, we may consider simulating a broader variety of anomaly types. Specifically, simulating non-symmetric or non-uniform anomalies could better capture the heterogeneity observed in dementia. Additionally, simulating smaller anomalies, which may challenge the detection using a VAE, would further enrich the evaluation.

Another potential limitation of the proposed evaluation framework is that the evaluation only relies on the difference between input and reconstructed images, i.e., the residual. The use of an anomaly mask that could be compared with the ground truth using a metric such as the dice score could be a great improvement. A simple solution may be thresholding the difference maps, or use Z-scores to attenuate the reconstruction noise and accentuate the anomaly \citep{solal:hal-04291561}. In addition, the only parameter for the simulation of anomalies is the degree of hypometabolism. However, establishing a correlation between this parameter and the progression or severity of the disease, as measured by cognitive scores (such as the MMSE or CDR), or the time elapsed before the first symptoms, poses a challenge. In other words, interpreting the intensity of the simulated anomalies in relation to the patient's cognitive status is not straightforward.

Choosing a VAE as generative model allowed us to perform analyses in the latent space. In particular, we used it to explain how the VAE behaves and interpret some of the results. We first observed that the latent representation of a same patient is always very close in the latent space, and almost identical between the simulated and the original PET scans (Figure~\ref{fig:latent_pca}). More globally we showed that the VAE encoder is able to map the complex data distribution to a simple multivariate Gaussian distribution of lower dimension. This explains why, for small deviations from the healthy image distribution (anomalies that we simulated), the model is able to reconstruct an image that is plausible (Figure~\ref{fig:results}), that corresponds to the patient under investigation, and seems to be pseudo-healthy. This can be explained by disentangling the functioning of the encoder and the decoder: the encoder catches the image structural information that is specific to the subject, and the decoder, given a latent representation $\textbf{z}$, can only reconstruct healthy looking images because it is what it has been trained to do. This is not straightforward and the model could have other behaviors. Indeed, we identify three different scenarios: 
\begin{itemize}
    \item the model reconstructs the identity, meaning that a healthy image has a healthy reconstruction and an abnormal image is reconstructed with its anomalies,
    \item the model does not reconstruct abnormal images at all as it has never seen some, which would be a behavior similar to out-of-distribution detection,
    \item the model reconstructs pseudo-healthy images since it could learn well the healthy image distribution, which seems to be the case.
\end{itemize}

Combining the latent space analysis with our simulation framework shows that the model has a similar behavior for the different kinds of anomalies (Table~\ref{tab:simulateddiseases}) and that the VAE can generalize well.

Our experiments on the latent space show that the encoder is working as we can expect. To improve the the quality of the reconstructed images, a first simple step would be to use a more powerful generator \citep{duquenne2022t}. If given a latent representation $\mathbf{z}$ the model can reconstruct perfectly the image $\mathbf{x}$, then we could detect anomalies with a high accuracy. Numerous research to improve the VAE framework has been performed in the computer vision literature \citep{chadebec2022pythae}. In a recent work, we benchmarked fifteen of them for the pseudo-healthy synthesis of brain FDG PET, without being able to improve to reconstruction quality compared with that of the vanilla VAE \citep{hassanaly2023benchmark}. 
We can imagine that combining a VAE with a diffusion model as done in \cite{pandey2022diffusevae} might be a good solution to improve the decoder. It showed great result on 2D images and future work could consist in comparing and evaluating such approach on a 3D task on all the different aspects we enunciated.

The proposed validation procedure is applicable outside of the use case presented here. Most of the code that we used is available in ClinicaDL \citep{thibeau-sutreClinicaDLOpensourceDeep2022}, an open-source software that is developed for reproducibility of deep learning studies in neuroimaging. Pipelines are available to perform the following steps:
\begin{itemize}
    \item selecting subjects from a neuroimaging dataset,
    \item rigorously separating data into independent training, validation and testing sets,
    \item easily training a VAE on neuroimages,
    \item constructing new test sets by generating simulated data using the proposed method,
    \item running tests to evaluate models.
\end{itemize}
Moreover, all the preprocessing pipelines are also available in Clinica \citep{clinica}, an open-source software for reproducible processing of neuroimaging datasets and multi-modal neuroscience studies. Clinica has been used to:
\begin{itemize}
    \item curate and organize the ADNI dataset following a community standard, namely the brain imaging data structure (BIDS) \citep{gorgolewski2016brain},
    \item perform linear registration and intensity normalization of the FDG PET scans.
\end{itemize}
Finally, all the code for training and evaluating the model is available on a Github repository: \url{https://github.com/ravih18/UAD_evaluation_framework}; and is tagged on Zenodo under the following DOI: \url{https://zenodo.org/doi/10.5281/zenodo.10568859}. % More details can be found in Appendix~\ref{}.}

\section{Conclusion}

We presented an extensive evaluation procedure of pseudo-healthy reconstruction for unsupervised anomaly detection in the case where ground truths are not available. It consists in different steps that are: the measurement of the reconstruction error on images from healthy subjects, the use of a simulation framework to create pairs of healthy and diseased images; the introduction of a metric to measure the healthiness of images when using the simulation framework; the use of a brain atlas to detect anomalies by comparing the input and the reconstructed images using the simulation framework and the real pathological images from ADNI dataset. 

This procedure has been applied to a 3D VAE that is suited to detect anomalies due to dementia on brain FDG PET. The VAE has been trained to reconstruct healthy-looking images using images of healthy subjects. We saw that the model can indeed reconstruct subject-specific pseudo-healthy images and can help to detect anomalies. We also validated the model ability to detect anomalies of different intensities, shapes and locations. However, the performance could be increased by improving the quality of the reconstruction.

We also exploited the latent space properties to understand the VAE behavior and interpret the results. We saw that the model can encode very similar latent representations for different images of a same subject (different sessions, or simulated images), and more generally, that the latent distribution represents well the image distribution. This is the expected behavior for the encoder. That means that if we want to improve the quality of the reconstruction, we would have to use a better decoder.

%%%%%%%%%%%%%%%%%%%%%%%%%%%%%%%%%%%%%%%%%%%%%%%%%%%%%%%%%%%%%%%%%%%%%%%
% Mandatory Sections. Please complete, especially for final publication
%%%%%%%%%%%%%%%%%%%%%%%%%%%%%%%%%%%%%%%%%%%%%%%%%%%%%%%%%%%%%%%%%%%%%%%

% Acknowledgements.
% Please include any funding, intellectual contributions not included in the authorship, and any other acknowledgements.
\acks{The research leading to these results has received funding from the French government under management of Agence Nationale de la Recherche as part of the "Investissements d'avenir" program, reference ANR-19-P3IA-0001 (PRAIRIE 3IA Institute) and reference ANR-10-IAIHU-06 (Agence Nationale de la Recherche-10-IA Institut Hospitalo-Universitaire-6).

This work was granted access to the HPC resources of IDRIS under the allocation AD011011648 made by GENCI (Grand Equipement National de Calcul Intensif).

Data collection and sharing for this project was funded by the Alzheimer's Disease Neuroimaging Initiative (ADNI) (National Institutes of Health Grant U01 AG024904) and DOD ADNI (Department of Defense award number W81XWH-12-2-0012). ADNI is funded by the National Institute on Aging, the National Institute of Biomedical Imaging and Bioengineering, and through generous contributions from the following: AbbVie, Alzheimer’s Association; Alzheimer’s Drug Discovery Foundation; Araclon Biotech; BioClinica, Inc.; Biogen; Bristol-Myers Squibb Company; CereSpir, Inc.; Cogstate; Eisai Inc.; Elan Pharmaceuticals, Inc.; Eli Lilly and Company; EuroImmun; F. Hoffmann-La Roche Ltd and its affiliated company Genentech, Inc.; Fujirebio; GE Healthcare; IXICO Ltd.; Janssen Alzheimer Immunotherapy Research \& Development, LLC.; Johnson \& Johnson Pharmaceutical Research \& Development LLC.; Lumosity; Lundbeck; Merck \& Co., Inc.; Meso Scale Diagnostics, LLC.; NeuroRx Research; Neurotrack Technologies; Novartis Pharmaceuticals Corporation; Pfizer Inc.; Piramal Imaging; Servier; Takeda Pharmaceutical Company; and Transition Therapeutics. The Canadian Institutes of Health Research is providing funds to support ADNI clinical sites in Canada. Private sector contributions are facilitated by the Foundation for the National Institutes of Health (\url{www.fnih.org}). The grantee organization is the Northern California Institute for Research and Education, and the study is coordinated by the Alzheimer’s Therapeutic Research Institute at the University of Southern California. ADNI data are disseminated by the Laboratory for Neuro Imaging at the University of Southern California.}

% Ethical Standards.
% Please edit with the appropriate ethics considerations for your work. Include any pertinent IRB information, etc.
%
% Please note that the submission requirements included:
% The work presented must follow appropriate ethical standards in conducting research and writing the manuscript, following all applicable laws and regulations regarding treatment of animals or human subjects.
\ethics{The work follows appropriate ethical standards in conducting research and writing the manuscript, following all applicable laws and regulations regarding treatment of animals or human subjects.}

% Conflict of Interest
% Declaration of possible conflicts of interest: Authors must disclose any financial, organisational, commercial or personal conflicts of interest that might bias their work.
% If no conflicts, please say "We declare we don't have conflicts of interest."
% \coi{The conflicts of interest have not been entered yet.}
\coi{We declare we do not have conflicts of interest.}

\bibliography{references}

% Manual newpage inserted to improve layout of sample file - not
% needed in general before appendices.
% \newpage

% Appendix is optional
\clearpage
\appendix

\renewcommand{\theHsection}{A\Alph{section}}

\section{Reconstruction metrics}
\label{sec:recon_metrics}

\paragraph{Mean squared error}

The MSE is the mean of the square of the difference between the true pixel ($X_{i}$) and the reconstructed pixel ($\widehat{X_{i}}$)
\begin{equation}
    MSE = \frac{1}{n} \sum_{i=1}^n \left( X_{i} - \widehat{X_{i}} \right)^{2}  \enspace .
\end{equation}
A low MSE means that the images are close to each other. They are identical if the MSE is 0.

\paragraph{Peak signal-to-noise ratio}

The PNSR is a function of the MSE and allows for comparing images encoded with different dynamic ranges
\begin{equation}
    PSNR =  10 \log_{10} \left( \frac{MAX^{2}}{MSE} \right)  \enspace ,
\end{equation}
with $MAX$ the maximum possible value of the image. We can see that if the images are similar, the MSE is close to 0, and the PSNR tends toward $+\infty$.

\paragraph{Structural similarity}

The SSIM is a weighted combination of three comparison measurements: the luminance $l$, the contrast $c$ and the structure $s$ \citep{wang2004image}
\begin{equation*}
    SSIM = l(X, \widehat{X})^{\alpha} \cdot c(X, \widehat{X})^{\beta} \cdot s(X, \widehat{X})^{\gamma} \enspace ,
\end{equation*}
with $\alpha$, $\beta$ and $\gamma$ the weights assigned to each measurement. If we set them all to 1, we obtain the following formula \citep{wang2004image}:
\begin{equation}
    SSIM = \frac{(2\mu_{X}\mu_{\widehat{X}} + c_1)(2\sigma_{X\widehat{X}} + c_2)}{(\mu_{X}^2 + \mu_{\widehat{X}}^2 + c_1)(\sigma_{X}^2 + \sigma_{\widehat{X}}^2 + c_2)} \enspace ,
\end{equation}
where:
\begin{itemize}
    \item $\mu_{X}$ and $\mu_{\widehat{X}}$ are the means of the true and reconstructed image respectively,
    \item $\sigma_{X}$ and $\sigma_{\widehat{X}}$ are the standard deviations of the true and reconstructed image respectively,
    \item $c_1$ and $c_2$ are positive constants to stabilize the division. Typical values are $c_1 = 0.01$ and $c_2 = 0.03$.
\end{itemize}

The SSIM ranges between 0 and 1.

\paragraph{Multi-scale structural similarity} 

The MS-SSIM is similar to the SSIM: it is a weighted combination of the luminance $l$, the contrast $c$ and the structure $s$ computed at different scales of the image \citep{wang2003multiscale}. To do so, we iteratively apply $M$ times a 3D average pooling, which is a low pass filter, to down-sample the image by a factor of two.

\begin{equation}
    MS\textnormal{-}SSIM = l_{M}(X, \widehat{X})^{\alpha_{M}} \cdot \prod_{j=1}^{M} c_{j}(X, \widehat{X})^{\beta_{j}} \cdot s_{j}(X, \widehat{X})^{\gamma_{j}} \enspace ,
\end{equation}
with  $l_{j}$, $c_{j}$ and $s_{j}$ being respectively the luminance, the contrast and the structure between $X$ and $\hat{X}$ at the scale $j$. We choose $M = 5$, $\alpha_{j} = \beta_{j} = \gamma_{j}$ and $\alpha_{j}, \beta_{j}, \gamma_{j}$ taking the following values $(0.0448, 0.2856, 0.3001, 0.2363, 0.1333)$ for $j\in[1, M]$ based on the values introduced in the original paper from \cite{wang2003multiscale}.

\section{Details of the simulation framework}
\label{sup:similation_framework}

We generated masks corresponding to six dementias: Alzheimer's disease (AD), behavioral variant frontotemporal dementia (bvFTD), logopenic variant primary progressive aphasia (lvPPA), semantic variant PPA (svPPA), nonfluent variant PPA (nfvPPA) and posterior cortical atrophy (PCA) based on the regions defined by \cite{Burgos2021AnomalyDetection} (Table \ref{tab:regions}).

\newpage

\begin{minipage}{\linewidth}
  \centering
  \captionof{table}{Regions associated with different dementia as defined in \cite{Burgos2021AnomalyDetection} and the masks used for hypo-metabolism simulation.}
  \begin{tabular}
      {|P{.15\linewidth}|p{.35\linewidth}|P{.40\linewidth}|} \hline Dementias & \centering Regions Associated & Masks \\
      \hline 
      
      \hfill \break Alzheimer's disease & 
        \begin{itemize}[noitemsep,topsep=0pt,leftmargin=*]
            \item \textbf{temporal lobe}, including the lateral and medial regions and temporal pole
            \item \textbf{parietal lobe}, including the superior and inferior regions
        \end{itemize}
        & 
      \includegraphics[align=t, width=\linewidth]{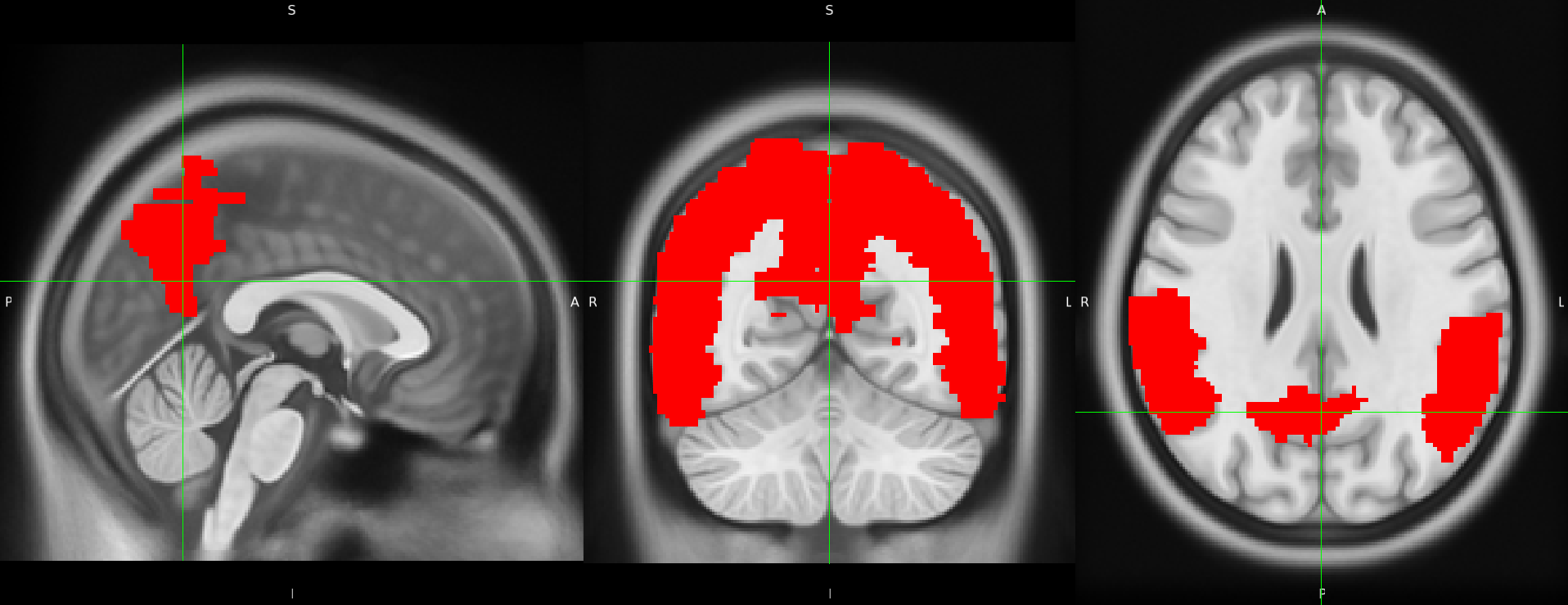} \\

      \hfill \break Behavioural variant frontotemporal dementia &
      \begin{itemize}[noitemsep,topsep=0pt,leftmargin=*]
           \item \textbf{orbitofrontal region}, comprising the anterior, posterior, medial, and lateral orbital gyri,
           \item \textbf{dorsolateral prefrontal region}, comprising the inferior, middle, and superior frontal gyri
           \item \textbf{ventromedial prefrontal region}, comprising the gyrus rectus, medial frontal cortex, subcallosal area, and superior frontal gyrus medial segment.
      \end{itemize}
      & 
      \includegraphics[align=t, width=\linewidth]{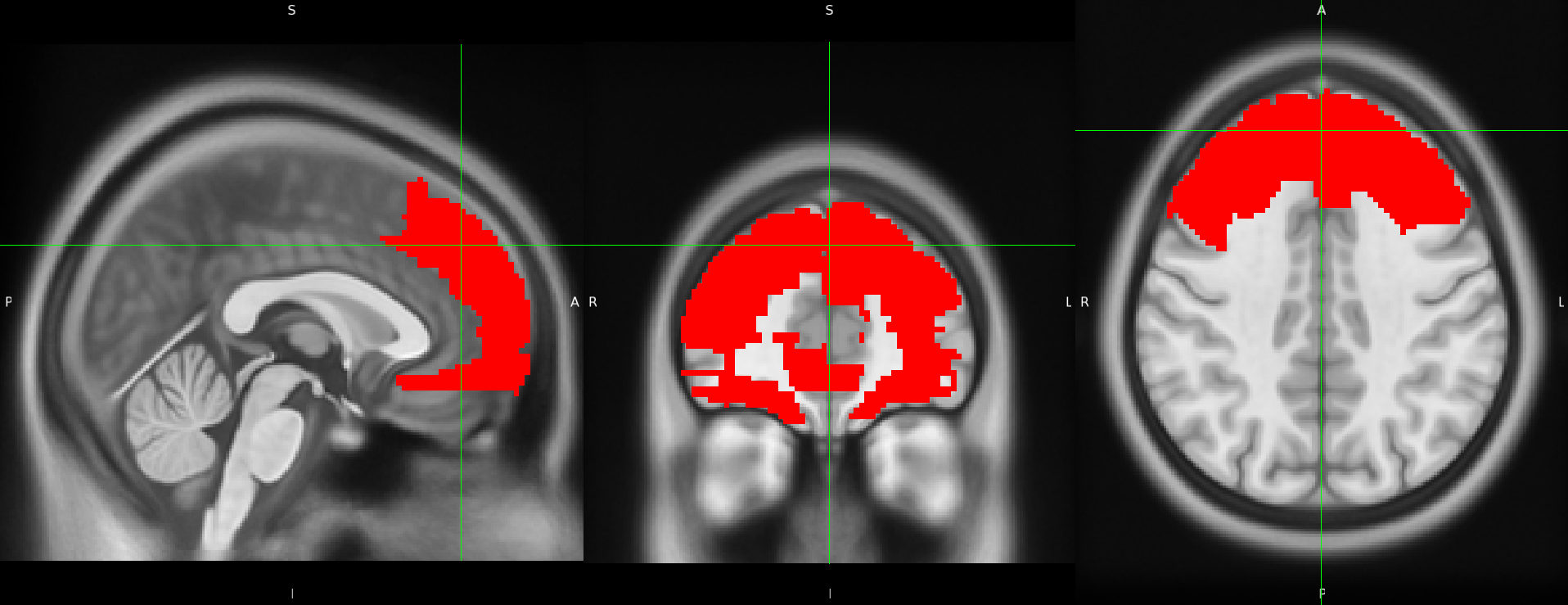} \\
      
      \hfill \break Posterior cortical atrophy &
      \begin{itemize}[noitemsep,topsep=0pt,leftmargin=*]
           \item \textbf{occipital region}, comprising the inferior, middle and superior occipital gyri.
      \end{itemize}
      & 
      \includegraphics[align=t, width=\linewidth]{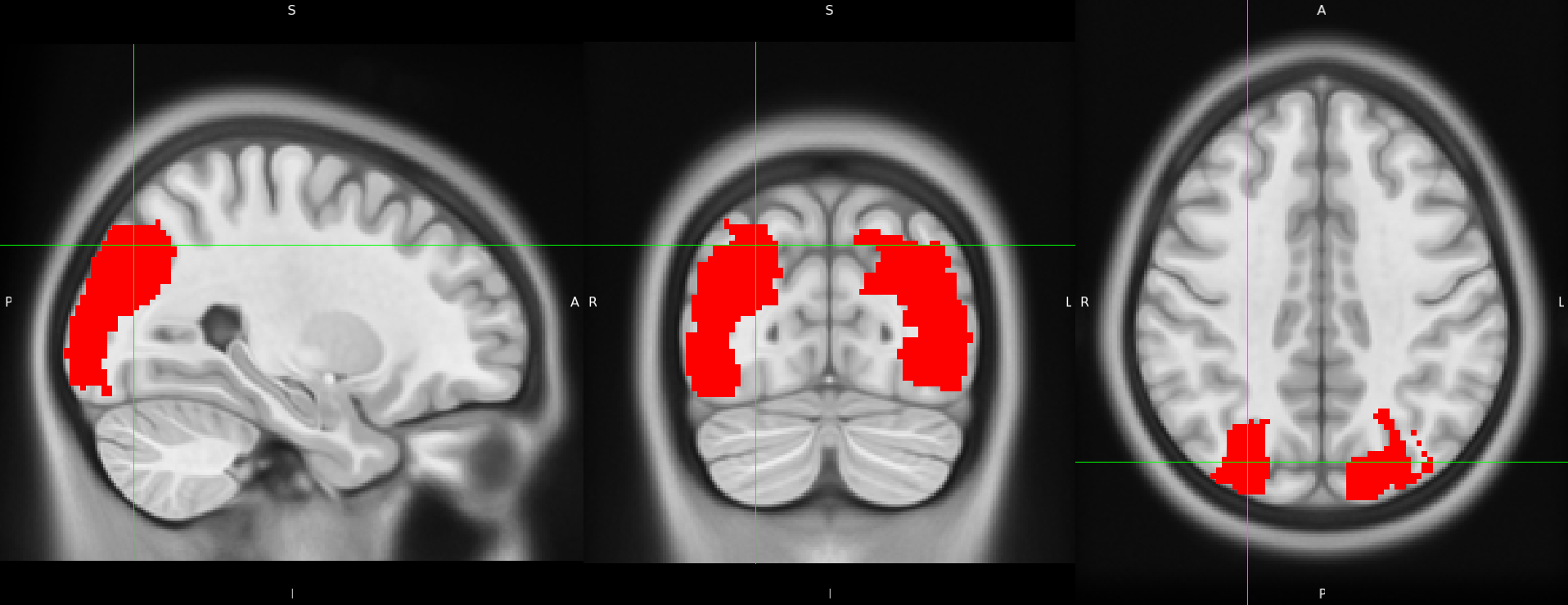} \\
      
      \hfill \break Logopenic variant primary progressive aphasia &
      \begin{itemize}[noitemsep,topsep=0pt,leftmargin=*]
           \item \textbf{tempoparietal region}, comprising the inferior parietal lobule, posterior middle and superior temporal gyri. 
      \end{itemize}
      & 
      \includegraphics[align=t, width=\linewidth]{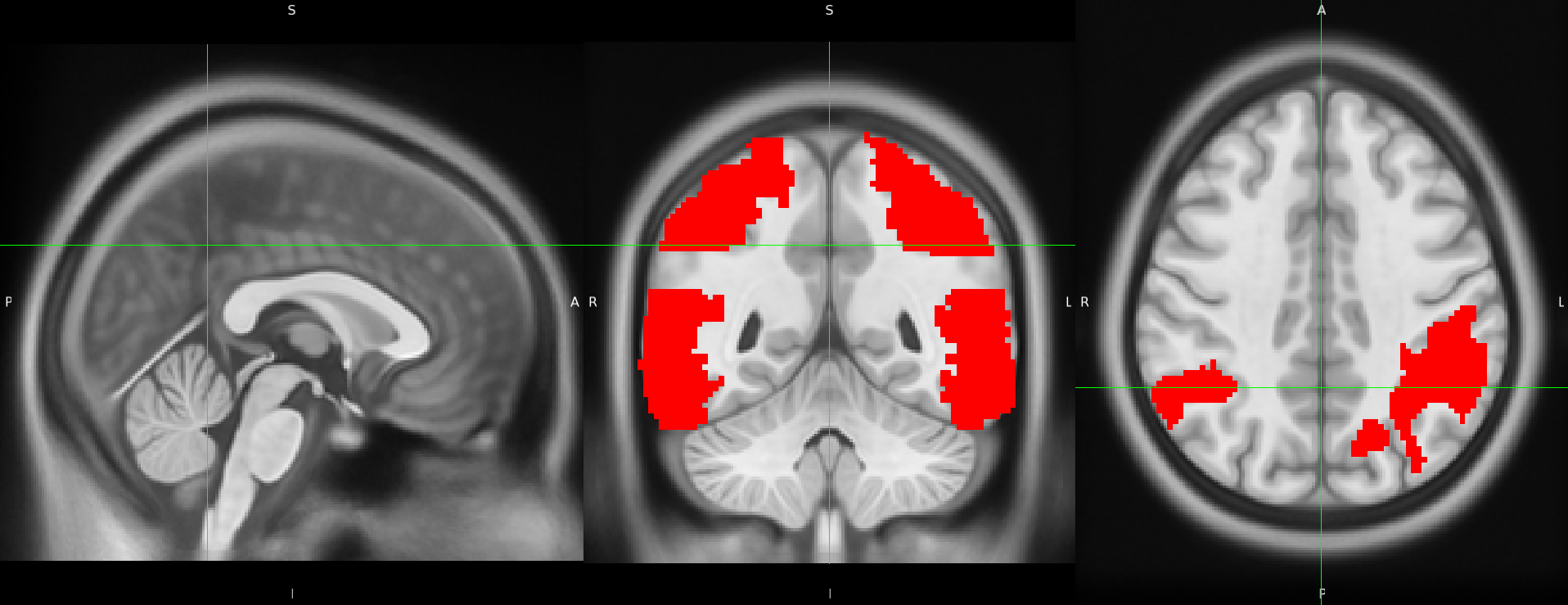} \\
      
      \hfill \break Nonfluent variant primary progressive aphasia &
      \begin{itemize}[noitemsep,topsep=0pt,leftmargin=*]
           \item \textbf{frontal region}, comprising the inferior frontal gyrus, precentral gyrus and anterior insula.
      \end{itemize}
      & 
      \includegraphics[align=t, width=\linewidth]{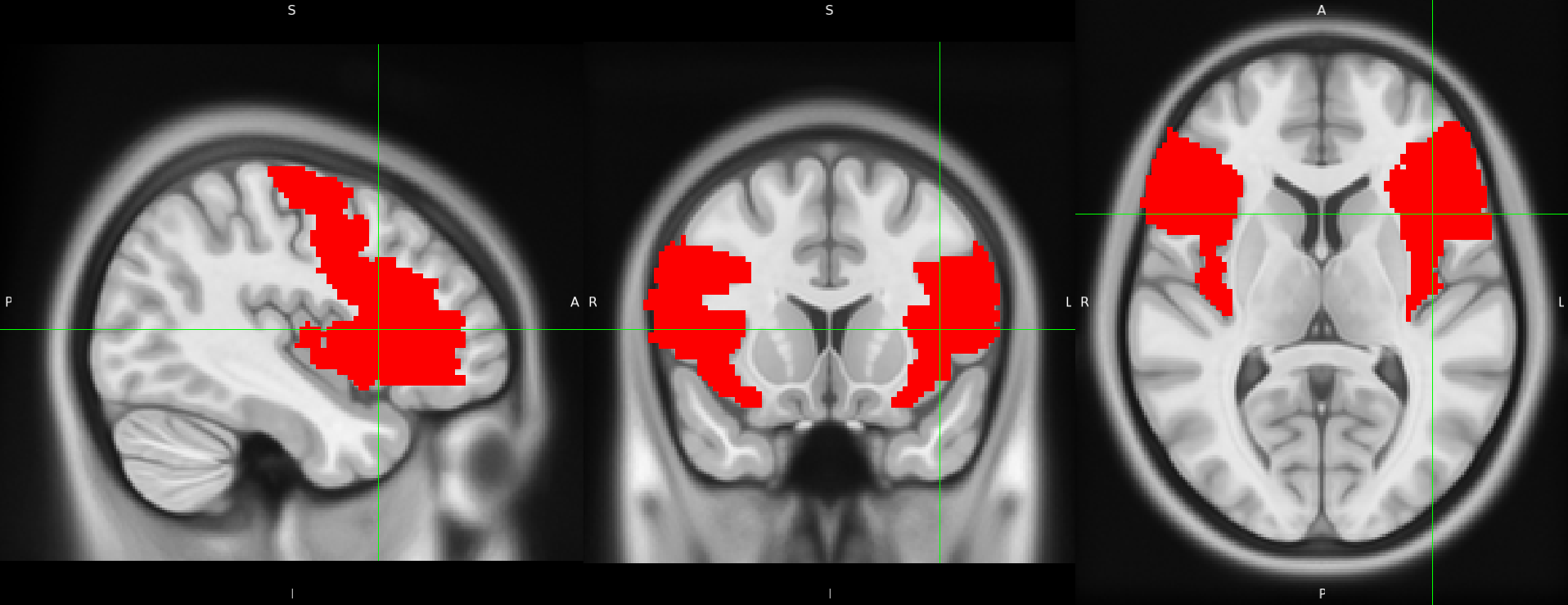} \\
      
      \hfill \break Semantic variant primary progressive aphasia &
      \begin{itemize}[noitemsep,topsep=0pt,leftmargin=*]
           \item \textbf{anterior temporal region}, comprising the hippocampus, amygdala and temporal pole.
      \end{itemize}
      & 
      \includegraphics[align=t, width=\linewidth]{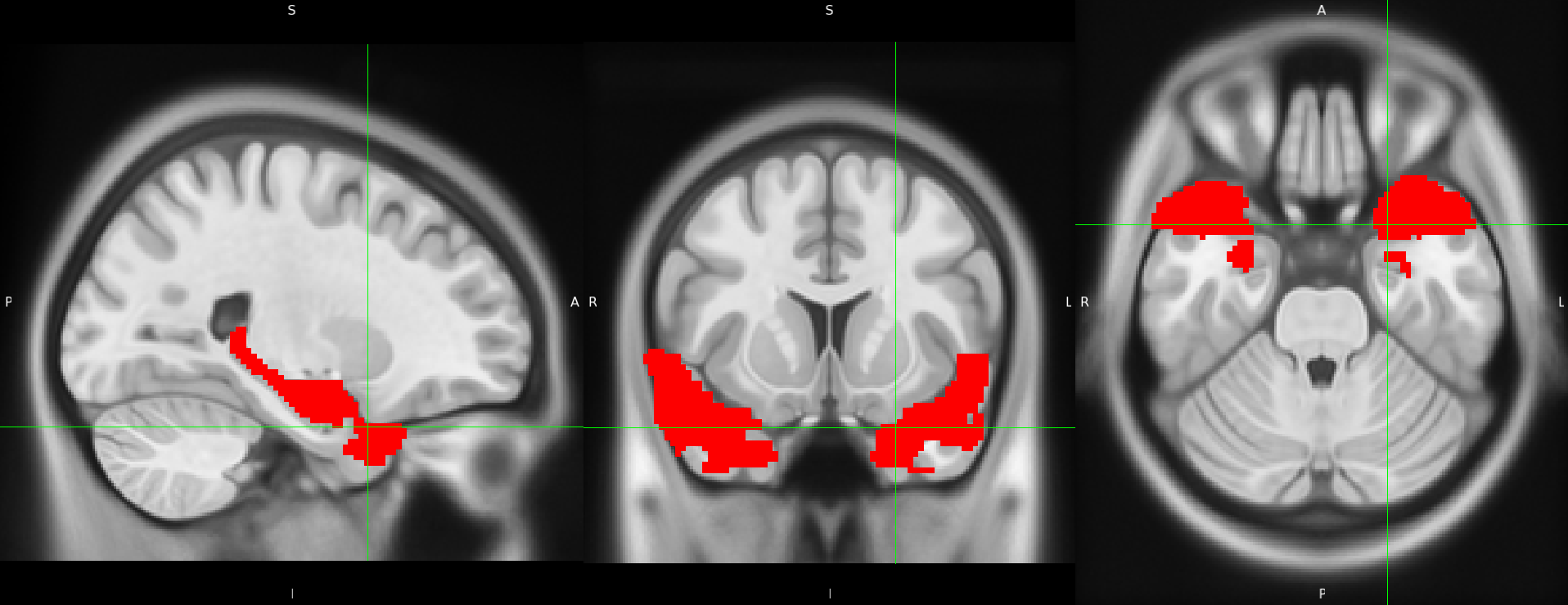} \\
      
      \hline
  \end{tabular}
  \label{tab:regions}
\end{minipage}
\newpage

\section{Linear mixed effect model}
\label{sec:lmm}
LMM is a statistical regression method used to analyze data that are dependent. It is particularly adapted to studies in which several observations are available per subject, such as in longitudinal studies. Here our different observations are the N closest images in the latent space (the tenth, the twentieth, the thirtieth and the fortieth closest images). The equation of the LMM is 
\begin{equation*}
    Y_{ij} = \beta_{0} + \beta_{1} X_{ij} + \gamma_{0i} + \gamma_{1i} X_{ij} + \epsilon_{ij} \enspace ,
\end{equation*}
where $(X_{ij}, Y_{ij})$ are respectively the distance in the latent space and the image space of the $j$th closest subject of subject i, $\beta_{0}$ and $\beta_{1}$ are the population effect parameters, $\gamma_{0i}$ and $\gamma_{1i}$ are the individual effect parameters and $\epsilon_{ij}$ the residual error. Each subject is modeled by a linear function of intercept $\beta_{0} + \gamma_{0i}$ and slope $\beta_{1} + \gamma_{1i}$. $\beta_1$ is the mean slope of the population and $\gamma_{1i}$ is the variance of each individual. We can then estimate the tendency of the evolution of the distance in the latent space with regards to the distance in the image space by observing the values of $\beta_{1} + \gamma_{1i}$.

\section{Data selection}
\label{sec:data_select}

In the ADNI database, there is a total of 3511 FDG PET scans from 1600 participants. This includes 554 cognitively normal (CN) subjects (1010 images) that we selected since UAD models are trained only on images from healthy subjects. We know that physiological changes can appear several years before the first clinical symptoms, so to ensure that images really correspond to a healthy brain, we kept only scans from subjects that are CN for at least three years after the session considered. We discarded 78 subjects (129 images) for whom diagnosis progresses to AD, 72 subjects (72 images) for whom there is a unique session (which is not enough to assess the reliability of the CN label) and 21 subject (49 images) for whom there are multiple conversions or regressions. We finally keep 383 stable CN subjects (760 images). 

There are also 560 AD patients (791 images). We removed 2 patients (2 images) with unstable AD diagnosis, 3 patients (3 images) of regressive AD, 189 patients (189 images) for whom there is a unique session, and 4 subjects that were already in the training set. In the end we keep the 362 baseline sessions of the remaining AD patients for testing purposes and discarded all the other images.

In addition, we ran both quality control procedures and discarded in total 30 images: 18 images from CN subjects and 9 images from AD patients after \texttt{t1-linear} quality control, and 3 images from CN subjects after \texttt{pet-linear} quality control. 

\section{Reconstruction metrics obtained on the validation set}
\label{sec:val}

\begin{minipage}{\linewidth}
    \captionof{table}{
        Reconstruction metrics on each validation set over the 6 folds.
    }
    \begin{center}
    \renewcommand{\arraystretch}{1.25}
    \begin{tabular}{l|c|cccc}
    Dataset                         & Fold  & MSE ($\times10^{-3}$) \textdownarrow	& PSNR \textuparrow	& SSIM \textuparrow	& MS-SSIM \textuparrow	\\ \hline 
    \multirow{6}{*}{Validation set}    & 0     & $2.043\pm0.899$	& $27.140\pm1.307$	& $0.862\pm0.036$	& $0.937\pm0.020$		\\
                                    & 1     & $1.988\pm0.634$	& $27.202\pm1.241$	& $0.866\pm0.025$	& $0.938\pm0.016$		\\
                                    & 2     & $1.865\pm0.766$	& $27.517\pm1.275$	& $0.877\pm0.031$	& $0.943\pm0.019$		\\
                                    & 3     & $1.891\pm0.584$	& $27.376\pm1.047$	& $0.872\pm0.037$	& $0.942\pm0.015$		\\
                                    & 4     & $1.907\pm0.654$	& $27.382\pm1.200$	& $0.868\pm0.027$	& $0.942\pm0.014$		\\
                                    & 5     & $2.005\pm1.098$	& $27.301\pm1.467$	& $0.868\pm0.035$	& $0.940\pm0.019$		\\
    \end{tabular}
    \end{center}
\label{tab:val_cn}
\end{minipage}

\begin{table}[!htb]
    \centering
    \caption{
        SSIM obtained on the validation set for all the splits.
    }
    \begin{tabular}{l|rrrrrr}
     & Split 0 & Split 1 & Split 2 & Split 3 & Split 4 & Split 5 \\ \hline
    mean & 0.861 & 0.865 & 0.876 & 0.871 & 0.868 & 0.867 \\
    std & 0.035 & 0.024 & 0.030 & 0.037 & 0.026 & 0.034 \\
    min & 0.698 & 0.801 & 0.741 & 0.66 & 0.795 & 0.736 \\
    25\% & 0.852 & 0.852 & 0.870 & 0.863 & 0.854 & 0.858 \\
    50\% & 0.870 & 0.872 & 0.884 & 0.881 & 0.870 & 0.873 \\
    75\% & 0.882 & 0.882 & 0.893 & 0.891 & 0.887 & 0.890 \\
    max & 0.912 & 0.905 & 0.923 & 0.907 & 0.908 & 0.907 \\
    \end{tabular}
    \label{tab:ssim_val}
\end{table}

\clearpage

\section{Examples of reconstructions obtained for healthy subjects and simulated hypometablic images}
\label{sup:supp_recon_cn}

\begin{minipage}{\linewidth}
    \centering
    \includegraphics[width=0.87\textwidth]{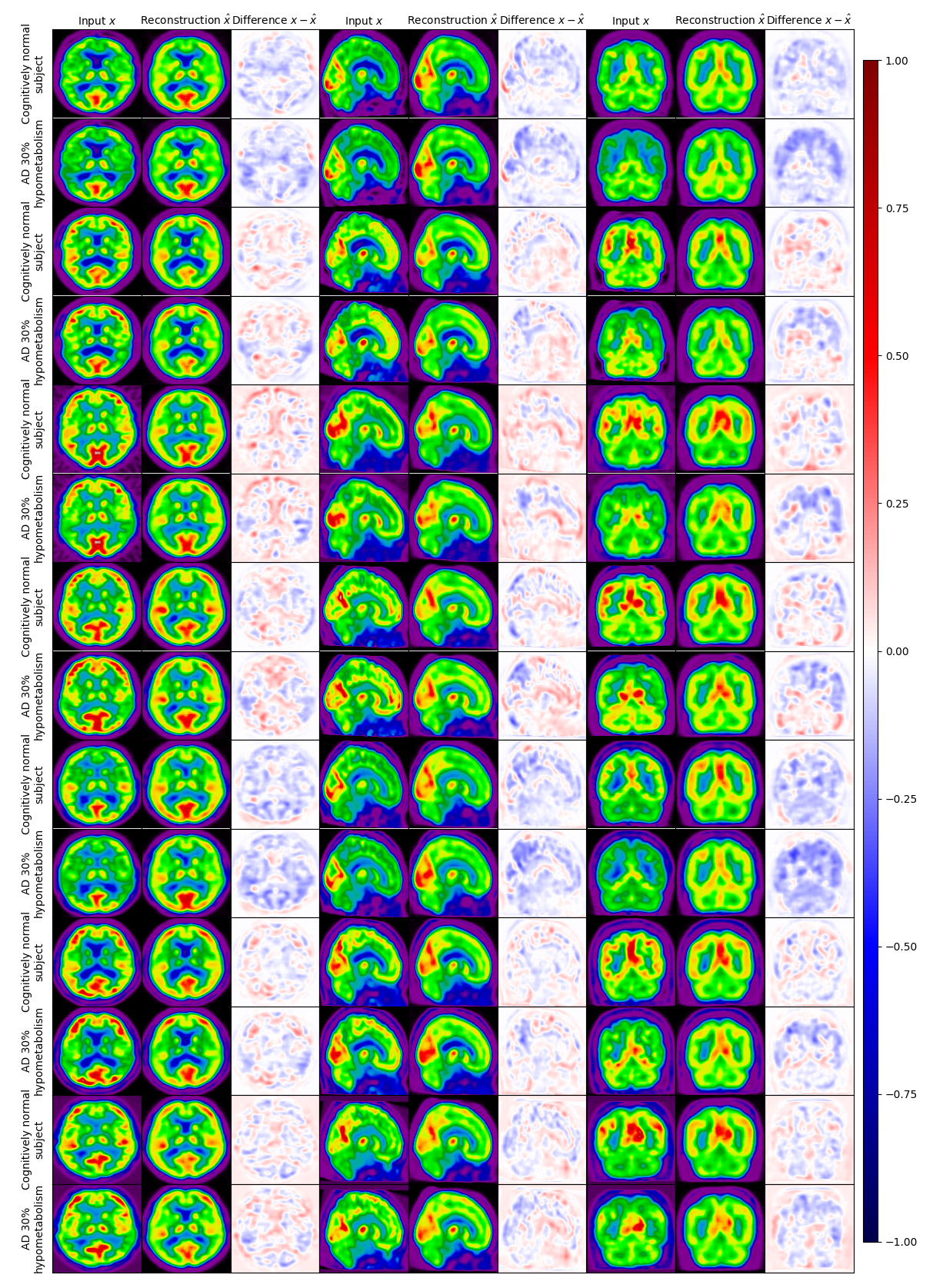}
    \captionof{figure}{
        Examples of reconstructions obtained from a real image of CN subjects (even rows) and the image simulating 30\% AD hypometabolism based on the same CN subject (odd rows). For each plane, the first image is the input, the second one the model's reconstruction and the third one the difference (input - reconstruction).
    }
    \label{fig:supp_recon_cn}
\end{minipage}

\section{Examples of reconstructions for AD patients}
\label{sup:recon_ad}

\begin{minipage}{\linewidth}
    \centering
    \includegraphics[width=0.9\textwidth]{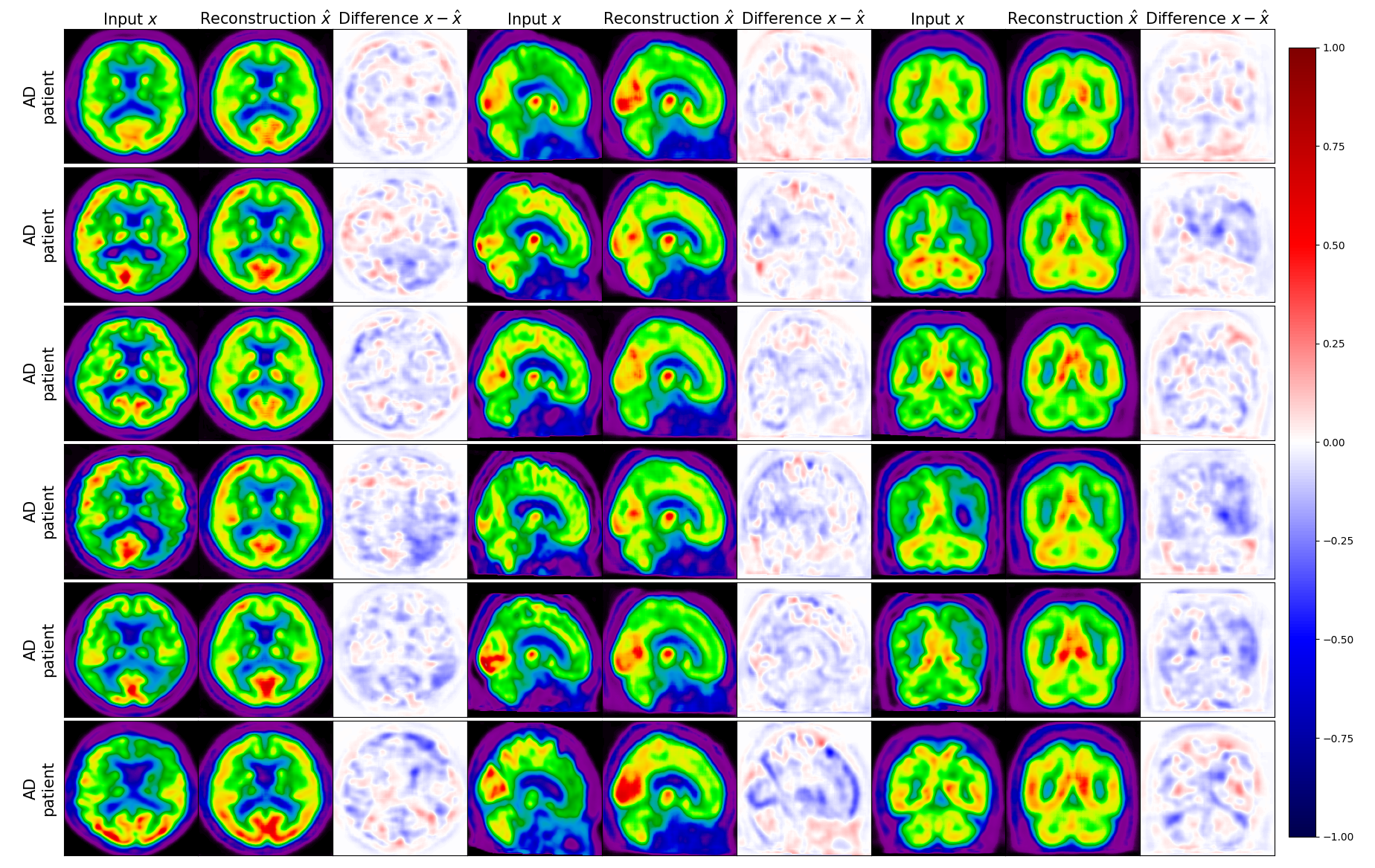}
    \captionof{figure}{
        Examples of reconstruction obtained from a real image of a AD patients. For each plane, the first image is the input, the second one the model's reconstruction and the third one the difference (input - reconstruction).
    }
    \label{fig:recon_ad}
\end{minipage}

\section{Detailed results of linear mixed effect models}
\label{sec:lmm_results}

\begin{table}[!htb]
    \centering
    \caption{
        Result of linear mixed effect models corresponding o Figure~\ref{fig:imgvslatent}. The top two rows correspond to the model fitted on the MSE against the Minkowski distance in the latent space. The bottom two rows correspond to the model fitted on the SSIM against the Minkowski distance in the latent space. "intercept" correspond to the intercept of the model and ``latent" to the slope. ``Coef." is the estimation of the value of the intercept or the latent, ``Std. Err." is the standard error on this value, $z$ is the z-score of this estimation, $P>|z|$ the p-value associated with this z-score and the last column is the confidence interval of the value.
    } 
    \renewcommand{\arraystretch}{1.25}
    \begin{tabular}{l l|c c c c c}
    & & Coef. & Std.Err. & $z$ & $P>|z|$ & [0.025 0.975]  \\
    \hline
    \multirow{2}{4em}{MSE ($\times10^{-3}$)}      & intercept & -7.806    & 0.583 & -13.397 & 0.000 & -8.948 -6.664   \\
                                    & latent    & 3.361     & 0.164 & 20.459 & 0.000 & 3.039  3.683 \\
    \hline
    \multirow{2}{4em}{SSIM}     & intercept & 0.932     & 0.009 & 106.935    & 0.000 & 0.915  0.949   \\
                                & latent    & -0.079    & 0.002 & -31.688   & 0.000 & -0.084 -0.074 \\
    \end{tabular}
    \label{tab:lmem}
\end{table}

\section{Unet results}
\label{sec:unet}

To demonstrate the value of the evaluation procedure, we trained a Unet with an architecture very similar to the VAE's one (we added skip connections and removed the probabilistic part in the latent space). The expected result of this experiment is that the model should be able to reconstruct good quality images by learning the identity function. Indeed, the model learning to reconstruct its inputs, it should probably use higher level skip-connections to minimize the reconstruction error. However, when trying to reconstruct images with anomalies, the model should also reconstruct an image similar to the input, instead of a pseudo-healthy version, as it did not learn any data distribution, but just an identity function.

We can observe in Table~\ref{tab:unet} that the reconstruction of the Unet is almost identical to the input image, with a SSIM of 0.99 on average. Compared to the VAE (Table~\ref{tab:result_cn}), the MSE is almost 100 times lower with the Unet. This means that the Unet is able to reconstruct images of high quality. However, we can see that the reconstructions are also similar to the input when the input is a simulated abnormal image, meaning that the model probably also reconstructs the anomalies.

\begin{table}[!htb]
    \centering
    \caption{
        Comparison of the reconstruction results obtained for Split 1 between the Unet and VAE.
    }
    \begin{tabular}{l|l|cccc}
        Model & Dataset & MSE ($\times10^{-3}$) \textdownarrow	& PSNR \textuparrow	& SSIM \textuparrow	& MS-SSIM \textuparrow	\\ \hline 
        \multirow{3}{*}{Unet}   & Test CN	& $0.024\pm0.005$ 	& $46.281\pm0.878$ 	& $0.990\pm0.001$ 	& $0.999\pm0.000$ 		\\
                                & Test AD	& $0.028\pm0.020$ 	& $45.864\pm1.555$ 	& $0.990\pm0.002$ 	& $0.999\pm0.000$ 		\\
                                & AD 30	& $0.024\pm0.006$ 	& $46.271\pm0.934$ 	& $0.990\pm0.001$ 	& $0.999\pm0.000$ 		\\ \hline
        \multirow{3}{*}{VAE}    & Test CN	& $1.815\pm0.649$ 	& $27.572\pm1.074$ 	& $0.878\pm0.026$ 	& $0.944\pm0.014$ 		\\
                                & Test AD	& $2.554\pm1.391$ 	& $26.272\pm1.560$ 	& $0.853\pm0.045$ 	& $0.928\pm0.025$ 		\\
                                & AD 30	& $2.345\pm0.639$ 	& $26.403\pm0.890$ 	& $0.869\pm0.027$ 	& $0.934\pm0.015$ 		\\

    \end{tabular}
    \label{tab:unet}
\end{table}

\begin{figure}[!htb]
    \centering
    \includegraphics[width=\textwidth]{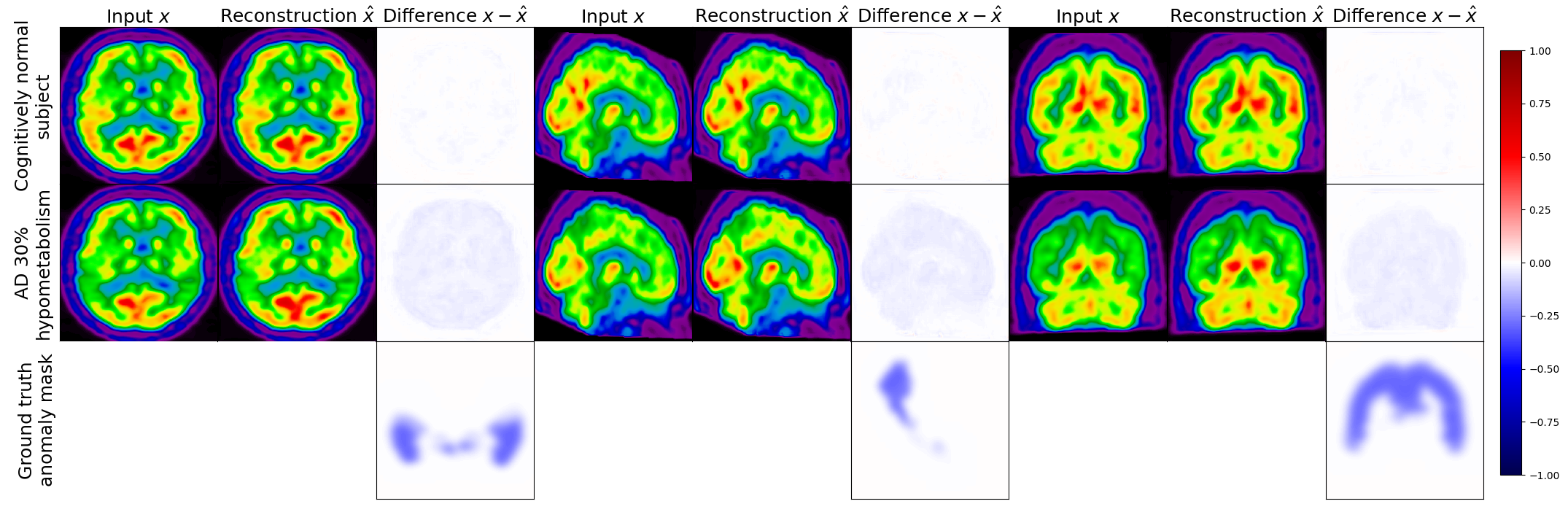}
    \caption{
        Example of results obtained with the Unet from a real image  of a CN subject (top row) and an image simulating AD hypometabolism based on the same CN subject (bottom row). For each plane, the first image is the input, the second one the model's reconstruction and the third one the difference (input - reconstruction). 
    }
    \label{fig:results_unet}
\end{figure}

This can be verified by observing directly the images in Figure~\ref{fig:results_unet}. We can see that the reconstruction is identical to the input, and the difference is null. We thus cannot detect anomalies.

\begin{figure}
    \centering
    \begin{subfigure}[b]{0.49\textwidth}
        \centering
        \includegraphics[width=\textwidth]{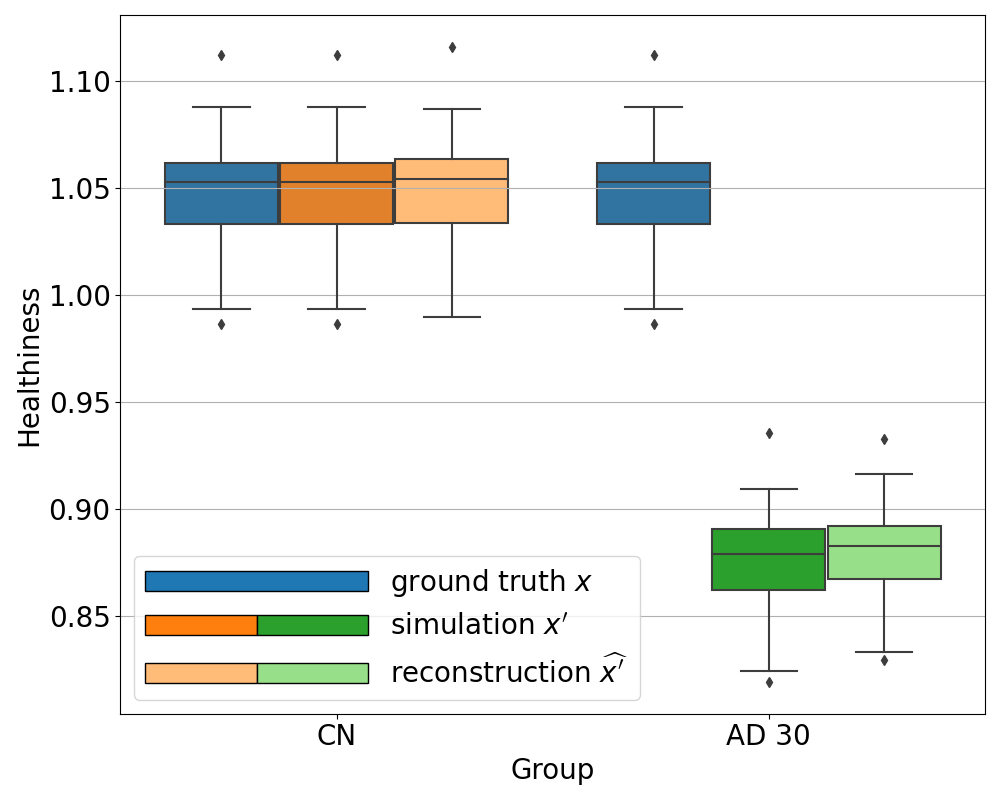}
        \caption{Unet}    
        \label{fig:healthiness_unet}
    \end{subfigure}
    \hfill
    \begin{subfigure}[b]{0.49\textwidth}  
        \centering 
        \includegraphics[width=\textwidth]{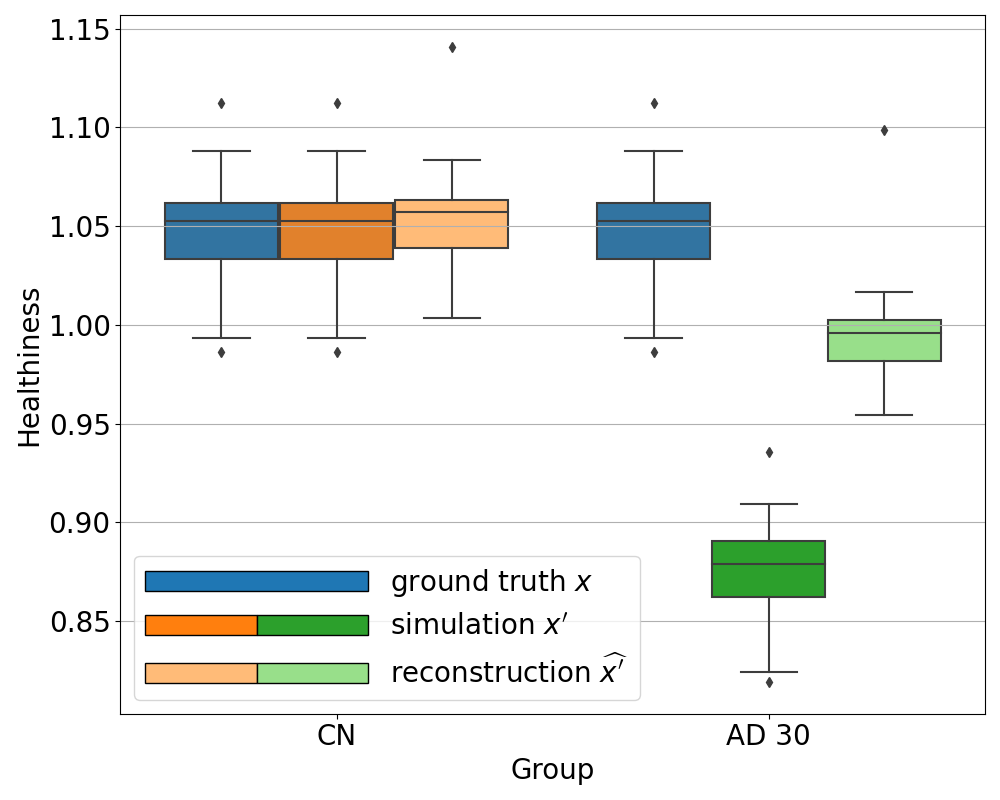}
        \caption{VAE} 
        \label{fig:healthiness_vae}
    \end{subfigure}
    \caption{
        Comparison of the distribution of the healthiness metric between the Unet and VAE for both CN and simulated AD subjects. For both models, the healthiness is constant (between 0.98 and 1.09) for images from healthy subjects of the test set (in blue), and the reconstruction is healthy as expected (in orange). For simulated images (AD 30), we can see that the healthiness is between 0.83 and 0.91 (dark green). However, the healthiness of the Unet reconstruction (light green) is the same (between 0.84 and 0.92), meaning that the reconstruction cannot be considered as pseudo-healthy. On the other hand, the healthiness of the VAE reconstruction, between 0.96 and 1.02, is higher than for the simulated image given in input. The VAE reconstruction can thus be considered as healthy.
    }
    \label{fig:healthiness_unet_vs_vae}
\end{figure}

This is confirmed when computing the healthiness metric defined in Section~\ref{sec:healthiness}. Even though we can see from the reconstruction metrics that the model is not able to reconstruct pseudo-healthy FDG PET images, this a limit case. In a more realistic scenario, reconstruction metrics and visual assessment of the images are not enough to estimate if a model is able to perform well. In Figure~\ref{fig:healthiness_unet}, we can see that the healthiness of the reconstructed image is equal to the one of the input image, meaning that the model could not reconstruct pseudo-healthy images.

\end{document}